\documentclass[aps,prb,twocolumn,superscriptaddress,showkeys,floatfix]{revtex4-2}
\usepackage[hypertexnames=true]{hyperref}
\usepackage{textcomp,gensymb}
\usepackage{graphicx}
\usepackage{lipsum}
\usepackage[utf8]{inputenc}
\usepackage{newunicodechar}
\usepackage[utf8]{inputenc}
\DeclareUnicodeCharacter{0308}{\"{}}   % combining diaeresis → \" accent
\newunicodechar{⁺}{\textsuperscript{+}}

\begin{document}
	
\title{Magnon Spin Current Modulation through Site-Specific Doping in a Compensated Iron Garnet
}

\author{Anna Merin Francis}
\author{P. B. S. Murthykrishnan}
\affiliation{Department of Physics, Indian Institute of Science Education and Research, Pune, India}

\author{Ratnamay Kolay}
\affiliation{School of Physics, Indian Institute of Science Education and Research, Thiruvananthapuram, India }
\author{Ramesh Nath}
\affiliation{School of Physics, Indian Institute of Science Education and Research, Thiruvananthapuram, India }
\author{Sunil Nair}
\affiliation{Department of Physics, Indian Institute of Science Education and Research, Pune, India}
%email{sunil@iiserpune.ac.in}
\date{\today}

\begin{abstract}

We report on the impact of manganese doping at the iron sites in Gadolinium Iron Garnet (GdIG, Gd$_{3}$Fe$_{5}$O$_{12}$), employing temperature-dependent spin Seebeck effect and ferromagnetic resonance measurements. Our findings reveal a clear shift in the magnetic compensation temperature \( T_{\text{comp}} \) in Mn-doped GdIG, with minimal changes observed in the magnetic and damping properties. Notably, the spin Seebeck signal strength was enhanced significantly with the Mn doping. This enhancement is attributed to an increased spin mixing conductance and modifications in the magnon spectra that strengthen exchange interactions, highlighting the material’s potential for room-temperature spintronic applications.

\end{abstract}

%\keywords{}

\maketitle


\section{Introduction}
Magnonics, the study of collective magnetic excitations, continues to advance the development of energy-efficient, high-performance magnetic storage and low-power information processing paradigms \cite{chumak2015magnon, pirro2021advances}. Rare-earth iron garnets have emerged as exemplary materials for pure spin current generation due to their intrinsically low magnetic damping, long spin-wave propagation lengths, and sub-micrometer spin-wave wavelengths \cite{PhysRevLett.115.096602}, with yttrium iron garnet (YIG) serving as the benchmark ferrimagnetic insulator for probing spin dynamics at the nanoscale \cite{chumak2015magnon, serga2010yig}. Since the seminal observation of the spin Seebeck effect (SSE) \cite{uchida2008observation}, a proliferation of studies have explored spin caloritronic \cite{bauer2012spin} phenomena across a broad spectrum of magnetic phase transitions, encompassing ferromagnetic \cite{jaworski2010observation}, ferrimagnetic \cite{uchida2010spin, adachi2013theory}, and antiferromagnetic \cite{wu2016antiferromagnetic, rezende2016theory} systems. Of particular interest are compensated ferrimagnets \cite{PhysRevB.87.014423}, which exhibit vanishing net magnetization at the magnetic compensation temperature ($T_{\text{comp}}$) due to the antiparallel alignment and mutual cancellation of distinct magnetic sublattices—typically comprising rare-earth and transition-metal ions—while retaining long-range magnetic order. By combining a tunable $T_{\text{comp}}$ and ferromagnet-like dynamic behavior with the ultrafast dynamics and negligible stray fields of antiferromagnets, compensated ferrimagnets provide a robust and versatile platform for the development of energy-efficient spintronic devices \cite{jungwirth2016antiferromagnetic, puebla2020spintronic,kim2022ferrimagnetic}.

\begin{figure}
	\includegraphics[width=1\columnwidth]{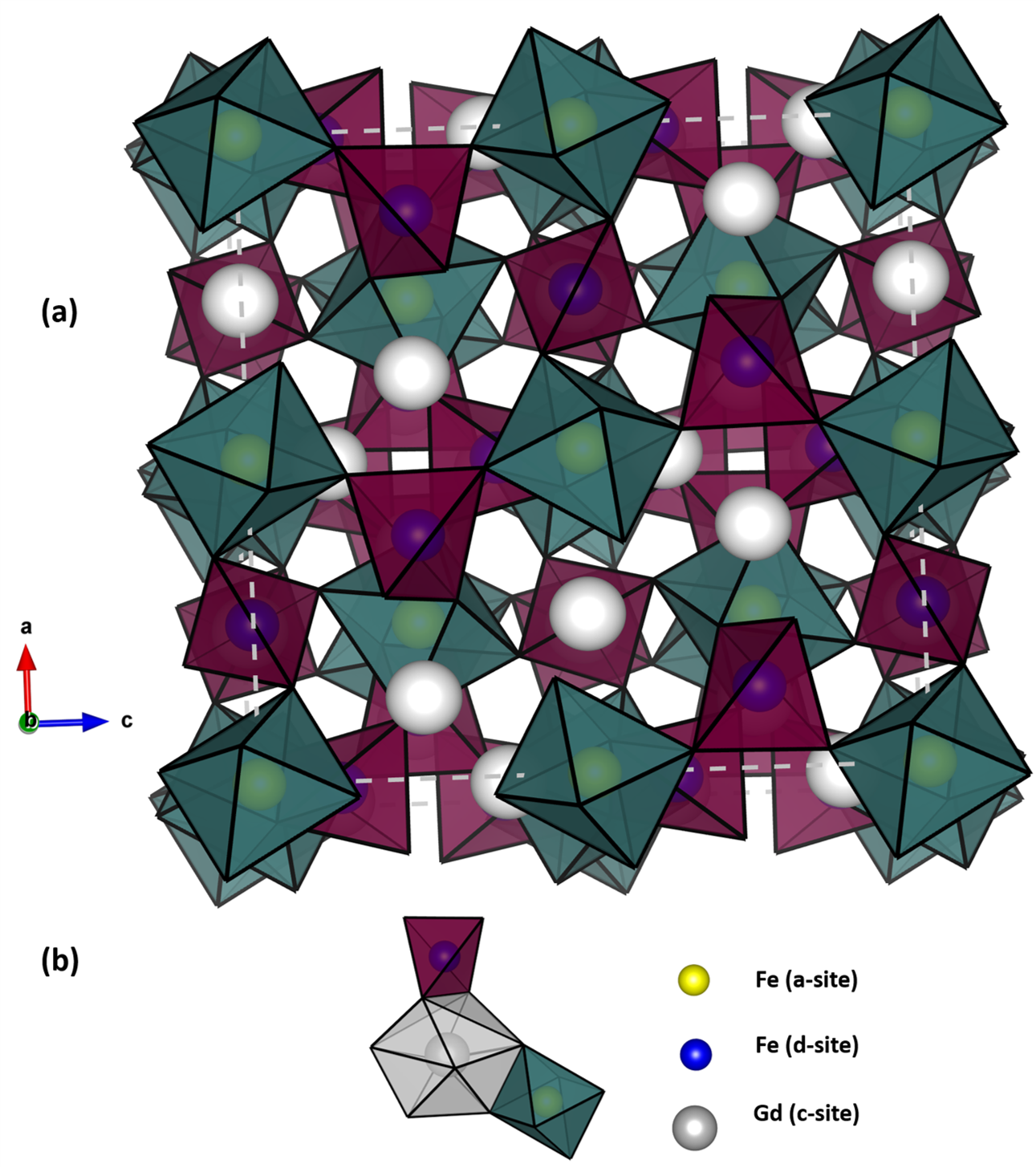}
	\caption{\footnotesize(a) Schematic cubic crystal structure of Gd$_{3}$Fe$_{5}$O$_{12}$. (b) Coordination environment of Fe and Gd ions, showing their surrounding oxygen polyhedra.}
	\label{struc}
\end{figure}

Gadolinium Iron Garnet (GdIG), a well-established compensated ferrimagnet \cite{geprags2016origin}, promises precise modulation of magnonic spin currents by adjusting its compensation temperature, surpassing YIG in versatility. The compensation point and magnetic properties of GdIG can be tailored through site-specific chemical doping at the Gd and Fe sublattices, as well as via strain engineering \cite{holzmann2022stress,liang2023thickness} techniques. GdIG thin films have also attracted considerable interest for their tunable, temperature-dependent perpendicular magnetic anisotropy (PMA) \cite{chanda2022scaling, chanda2024thermally, PhysRevB.96.224403}, influenced by magnetoelastic strain from substrate lattice mismatch. The presence of PMA around the \( T_{\text{comp}}\) positions GdIG as a promising candidate for high-density magnetic data storage applications.

\begin{figure*}
	\includegraphics[width=2.1\columnwidth]{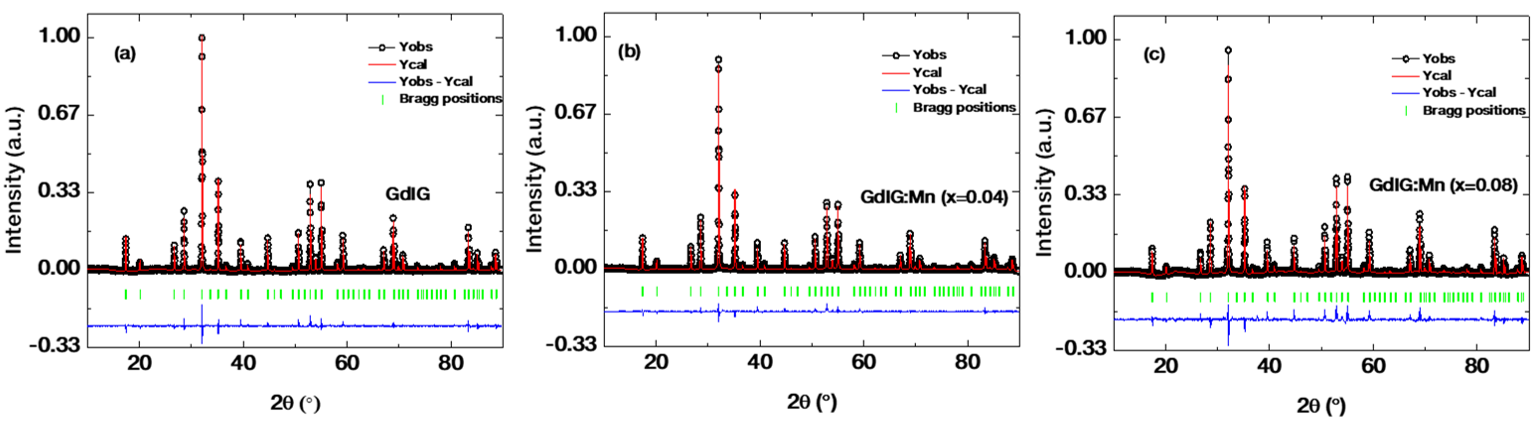}
\centering
    \caption
    {\footnotesize Rietveld refinement of the powder X-ray diffraction data of (a) GdIG, (b) GdIG:Mn (x=0.04), and (c) GdIG:Mn (x=0.08) measured at room temperature. All samples crystallize in the cubic symmetry with space group \textit{Ia{3}d}. The observed data are shown as black markers, while the red lines represent the calculated profiles. Vertical green ticks indicate the Bragg peak positions, and the blue line at the bottom shows the difference between the observed and calculated intensities.}
	\label{fig:three graphs} 
\end{figure*}

The crystal structure of GdIG, depicted in FIG. \ref{struc}, comprises a unit cell containing 12 trivalent Fe atoms in tetrahedral coordination (d-site), 8 Fe atoms in octahedral coordination (a-site), and 12 Gd atoms occupying dodecahedral (c-site) positions. The magnetism in GdIG predominantly stems from localized Fe$^{3+}$ moments, arising from an antiparallel ferrimagnetic alignment between  Fe$^{3+}$ ions occupying the a-site and d-site sublattices. A weak antiferromagnetic exchange between the Gd$^{3+}$ and Fe$^{3+}$ sublattices, combined with the strong temperature dependence of the Gd$^{3+}$ magnetic moment, gives rise to the $T_{\text{comp}}$ in GdIG. In this study, we explore the effect of Mn substitution at the Fe d-sites in GdIG. The rationale behind this approach is that by replacing d-site Fe with an element like Mn, which has a lower magnetic moment, we can effectively reduce the magnetic contribution from the d-site. This, in turn, could shift the \( T_{\text{comp}} \) closer to room temperature. We further investigate how Mn doping influences the magnon-mediated spin Seebeck effect, aiming to understand its impact on both magnetic compensation behavior and pure spin current generation. These characteristics can be leveraged in developing prototype spin-based devices \cite{wehmeyer2017thermal, finocchio2021perspectives}.

\section{Experimental Methods}

Polycrystalline samples of Gd$_{3}$Fe$_{5}$O$_{12}$ (GdIG) and Mn-doped GdIG, Gd$_{3}$Fe$_{5-x}$Mn$_x$O$_{12}$ with $x$ = 0.04 and 0.08 [denoted as GdIG:Mn (x=0.04) and GdIG:Mn (x=0.08)], were synthesized via the conventional solid-state reaction method using high-purity precursor powders: Fe$_{2}$O$_{3}$ (99.9\%), Gd$_{2}$O$_{3}$ (99.99\%), and Mn$_{2}$O$_{3}$ (99.99\%). The powders were weighed in stoichiometric amounts and ground into a homogenized mixture using a mortar and pestle. The mixing process was facilitated using ethanol (99.99\%), and the resulting mixture was subsequently compacted into pellets. The GdIG sample was sintered at 1200$^{o}$C for 10 hours in an air atmosphere inside a box furnace. A differential heating rate was applied in order to inhibit the formation of the perovskite phase GdFeO$_{3}$, which stabilizes around 800$^{o}$C to 1000$^{o}$C. The GdIG:Mn samples were sintered at 1200$^{o}$C in an oxygen-rich atmosphere.  Unlike GdIG, the Mn-doped samples required a two-step sintering process, involving an initial 10-hour sintering followed by an additional 24 hours to achieve a pure phase. An oxygen-rich atmosphere was employed in order to restrict the decomposition of the Mn$_{2}$O$_{3}$ into Mn$_{3}$O$_{4}$ at high temperatures. 

Laboratory X-Ray Diffraction measurements at room temperature were performed using a Bruker D8 Advance Diffractometer (CuK$_{\alpha}$ $\lambda$=1.5406\AA), confirming the sample to be single-phase. The Rietveld refinements were performed for room temperature XRD using \textit{Fullprof Suite} as shown in FIG.\ref{fig:three graphs}. Elemental compositions and their homogeneity were reconfirmed using an energy dispersive X-Ray spectrometer (Ziess Ultra Plus). The average stoichiometry determined from the statistical analysis of the EDS data confirmed the appropriate doping ratio in the synthesized samples. Temperature-dependent dc magnetization (M) in zero field cooling (ZFC) mode was measured using a vibrating sample magnetometer (VSM) attached to Physical Property Measurement System (PPMS, Quantum Design) in the temperature range $1.9\ \mathrm{K} \leq T \leq 380\ \mathrm{K}$ and magnetic field range $-0.5 \leq H \leq 0.5\ \mathrm{T}$.

The SSE measurements were conducted over a temperature range of 15–315K using a custom-built closed-cycle refrigerator (CCR) based apparatus, equipped with an electromagnet in the longitudinal SSE (LSSE) configuration \cite{PhysRevLett.124.017203}. This setup maintained a consistent 20K temperature gradient across the polycrystalline slab at an applied external magnetic field of up to 0.2T. The measured voltage corresponds to the average of signals taken at opposite magnetic field directions, where the background signal is canceled out through subtraction. For the SSE voltage detection, a 20nm platinum layer (Pt) was coated onto the finely polished surfaces of the polycrystalline slabs using DC magnetron sputtering. The temperature-dependent FMR measurements were performed using a home-built, broadband CryoFMR system in conjunction with a CCR and an electromagnet. The vector network analyzer (VNA) based measurements were conducted on 0.3mm thick polycrystalline slabs, positioned on a grounded coplanar waveguide (GCPW) fabricated using Rogers laminate. These measurements spanned 20–350K in an in-plane FMR geometry under an applied magnetic field of up to 0.6T. For each temperature, frequency-dependent (7–18GHz) magnetic field measurements were carried out to investigate the FMR absorption spectra.

\section{Results and Discussion}

The structural refinement confirms that all samples crystallized in the cubic \textit{Ia{3}d} symmetry. The Mn doping induces only minimal changes in the lattice parameters, with GdIG, GdIG:Mn (x=0.04) and GdIG:Mn (x=0.08) exhibiting unit cell dimensions of 12.4760(3)\text{\AA}, 12.4750(3)\text{\AA}, and 12.4750(2)\text{\AA}, respectively. This negligible variation is consistent with the close similarity in the ionic radii of Fe³⁺ and Mn³⁺. The site occupancy analysis reveals that Mn is partially incorporated into both the tetrahedral and octahedral Fe sublattices. In GdIG:Mn (x=0.04), Mn atoms exhibit a preferential occupancy of approximately 56\% at the tetrahedral sites compared to the octahedral ones. However, at a higher doping level (x=0.08), the Mn distribution becomes more balanced, with equal (50\%) occupancy at both tetrahedral and octahedral sites.
\begin{figure}
	\includegraphics[width=1\columnwidth]{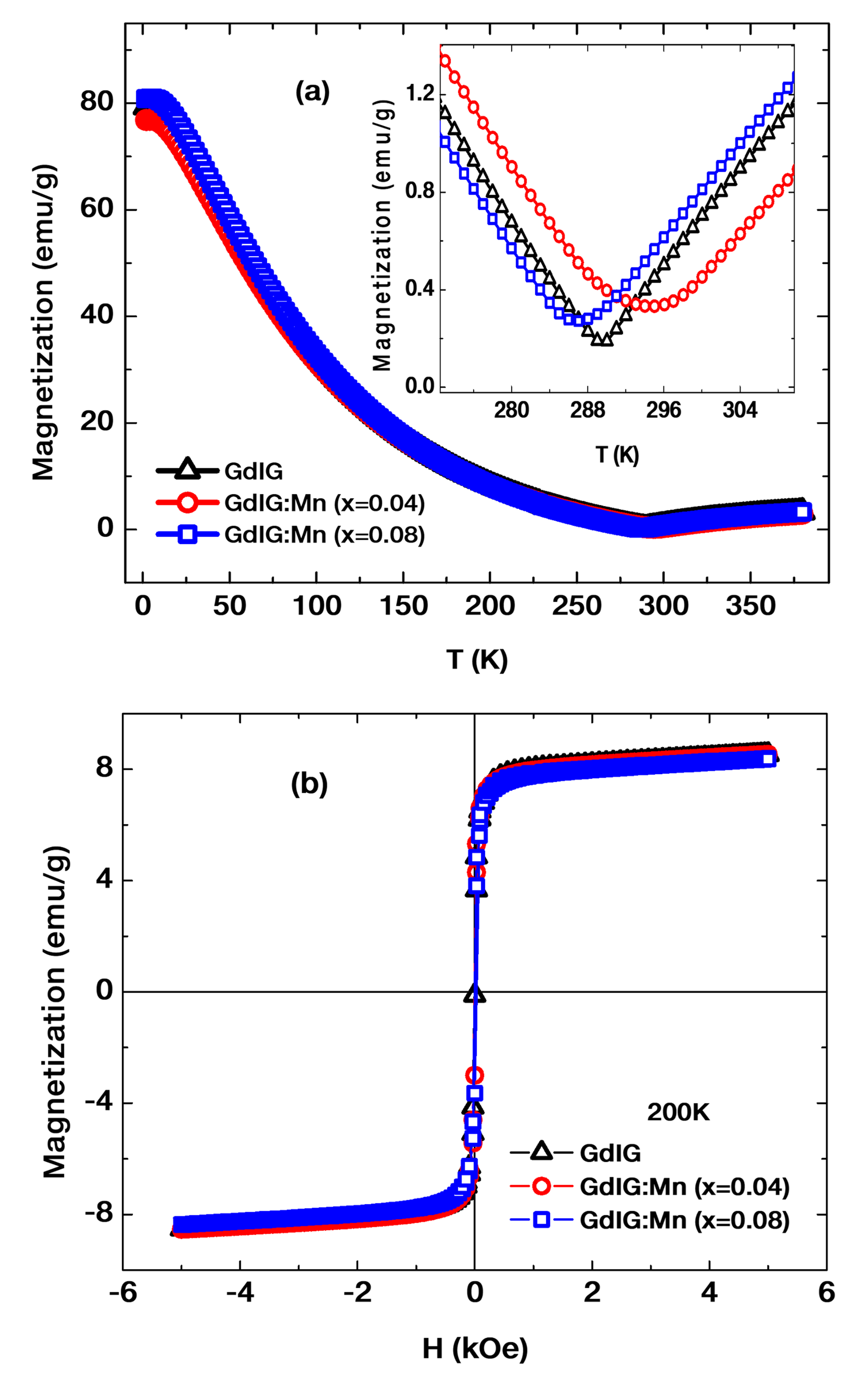}
\centering
    \caption
    {\footnotesize (a) Temperature dependence of magnetization measured under zero-field-cooled (ZFC) conditions at an applied magnetic field of 2kOe, with an inset showing an enlarged view of the magnetic compensation region. (b) Magnetic hysteresis loops recorded at 200K for polycrystalline slabs of GdIG, GdIG:Mn (x=0.04), and GdIG:Mn (x=0.08). A systematic shift in the $T_{\text{comp}}$ is noticed with Mn doping, while the overall magnetic profile remains similar. }
	\label{magnetic} 
\end{figure} 
Temperature-dependent magnetization measurements reveal a distinct shift in the magnetic compensation temperature ($T_{\text{comp}}$) with increasing Mn doping, while the overall net magnetization remains largely unaffected.  In GdIG, the $T_{\text{comp}}$ is observed at 290K; upon Mn doping, it rises to 297K in GdIG:Mn (x=0.04) and then decreases to 286K in GdIG:Mn (x=0.08). The observed trend underscores the sensitivity of $T_{\text{comp}}$ to site-specific magnetic contributions in GdIG. A reduction in tetrahedral-site magnetization increases the net Fe magnetic moment, which in turn leads to an increase in the $T_{\text{comp}}$. Conversely, a decrease in octahedral-site magnetization lowers the net Fe moment, thereby reducing $T_{\text{comp}}$. In GdIG:Mn (x=0.04), the higher concentration of Mn at the tetrahedral site results in an upward shift of $T_{\text{comp}}$ toward room temperature. In contrast, in GdIG\:Mn (x=0.08), where Mn is more evenly distributed between the octahedral and tetrahedral sites, $T_{\text{comp}}$ decreases slightly, falling below that of GdIG. The effective magnetic moments, estimated from d-orbital configurations and crystal field splitting for high-spin states—7/2 for Gd³⁺, and 5/2 for Fe³⁺ and 2 for Mn³⁺—in tetrahedral and octahedral coordination, yield approximately $19\,\mu_\mathrm{B}$ across all samples. The magnetic hysteresis measurements at 200K show negligible changes in coercivity ($H_{\text{C}}$), remanent magnetization ($M_{\text{R}}$), and saturation magnetization ($M_{\text{S}}$) across all compositions.
The temperature dependence of the magnetization in GdIG and GdIG:Mn slabs is shown in FIG.\ref{magnetic}.
\begin{figure*}
	\includegraphics[width=2.1\columnwidth]{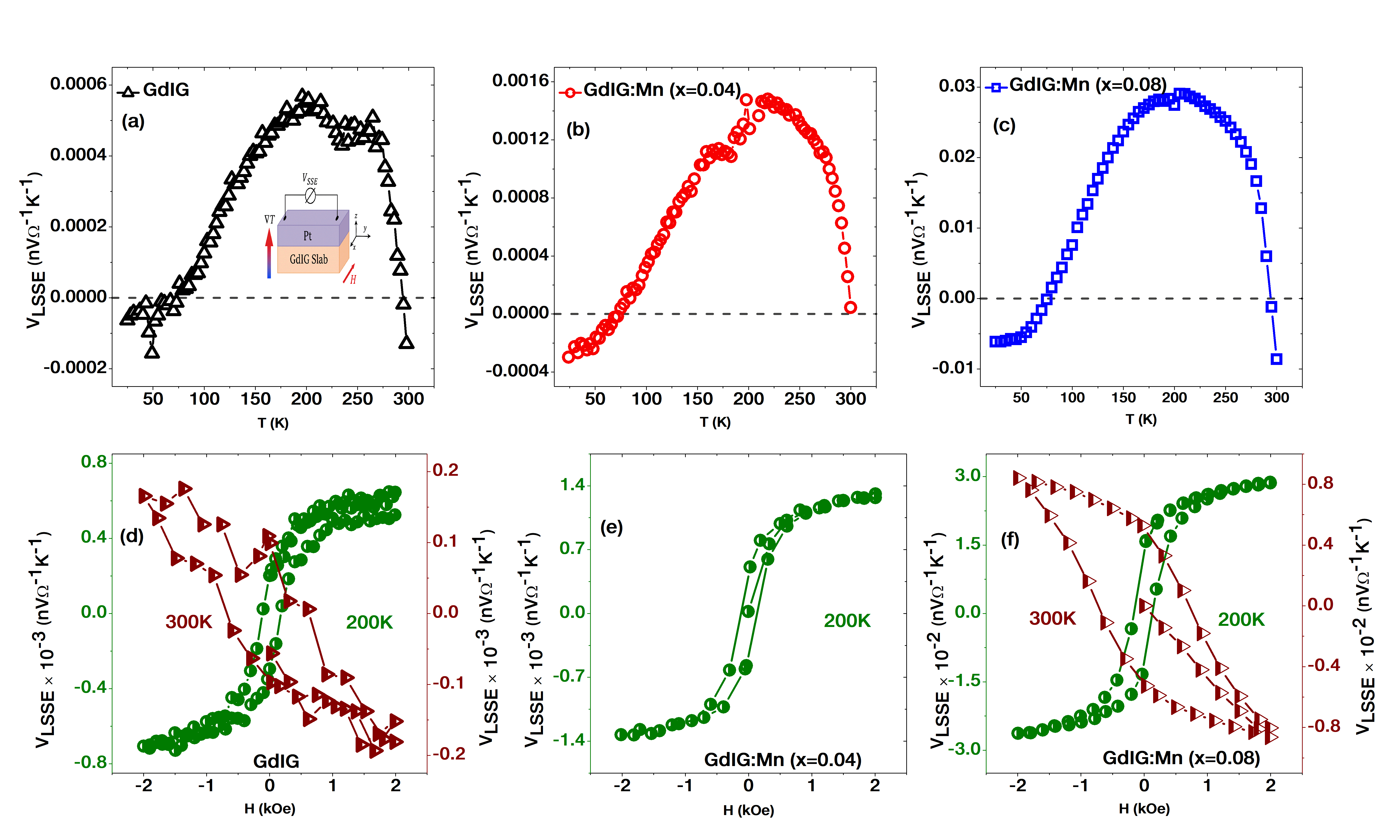}
\centering
    \caption
    {\footnotesize  Temperature-dependent SSE measurements on polycrystalline slabs of (a) GdIG, (b) GdIG:Mn (x=0.04), and (c) GdIG:Mn (x=0.08), measured under an applied magnetic field of 2kOe and a fixed temperature gradient of 20K. The schematic of the device used for longitudinal SSE measurements is shown in FIG. (a). Magnetic field dependent SSE measurements on (d) GdIG, (e) GdIG:Mn (x=0.04), and (f) GdIG:Mn (x=0.08), measured under a fixed temperature gradient of 20K, show polarity reversal above and below the $T_{\text{comp}}$. The magnetic field-dependent SSE plot for GdIG:Mn (x=0.04) at 300K could not be measured due to the proximity to $T_{\text{comp}}$ and hence vanishingly small signal strength.}
	\label{SSE} 
\end{figure*}

Temperature-dependent SSE measurements were performed on Pt-coated GdIG and GdIG:Mn slabs, with the resulting spin current detected via the Inverse Spin Hall Effect (ISHE) \cite{uchida2010observation}. As the measured voltages are affected by differences in effective resistance, the data are presented in a normalized form as follows:
\begin{equation}
\ V_{LSSE}={V_{dc}}t/RL{\Delta T} 
\end{equation}
where \(V_{dc} \) is the observed transverse voltage, t is the thickness of the polycrystalline slab, R and L are the corresponding resistance of the Pt-bar,  and the distance between contact probes, \({\Delta T}\) is the temperature gradient across the sample.\\  
The temperature-dependent SSE measurements reveal two distinct transition temperatures at which the SSE signal vanishes. The higher temperature transition \(T_{\text{sign1}} \) appears near 294K in both GdIG and GdIG:Mn (x=0.08), whereas in GdIG:Mn (x=0.04), it is observed above 300K. The corresponding temperature-dependent LSSE measurements for all three samples are shown in FIG.~\ref{SSE}. A second transition, occurring at a lower temperature \(T_{\text{sign2}} \), is consistently observed around 75K in all cases. The disappearance of the SSE signal at the $T_{\text{comp}}$ is attributed to the vanishing net magnetization of GdIG, resulting from the complete cancellation of the magnetizations of the Gd³⁺ and Fe³⁺ sublattices. As a result, the acoustic magnon mode ($\alpha$) softens, and consequently, the spin currents generated by the two sublattices cancel each other, as their magnons carry opposite spin polarizations. The reversal of the spin quantization axis at $T_{\text{comp}}$ results in the reversal of the SSE occurring at the \(T_{\text{sign1}} \) \cite{geprags2016origin, cramer2017magnon}.
At very low temperatures, thermal magnons from the optical magnon modes $\beta$ are largely suppressed due to the presence of a significant magnon gap. And the SSE signal below approximately 75K is primarily governed by $\alpha$ modes, which are predominantly associated with the Fe sublattices. These Fe-derived magnons couple efficiently to the Pt layer due to the strong spin mixing conductance \({g_{\uparrow\downarrow}}\) arising from the overlap between Fe 3$d$ electrons and Pt conduction electrons. As the temperature increases beyond 75 K, the magnon gap becomes thermally accessible, enabling activation of the $\beta$ modes, which are more closely linked to the Gd sublattice. These $\beta$ magnons carry spin angular momentum opposite to that of the $\beta$ and interact more weakly with Pt due to the localized nature of the Gd 4$f$ electrons and their lower spin mixing conductance. As their population increases with temperature, the $\beta$ magnons begin to dominate the spin current, leading to a reversal of the SSE voltage around 75K \cite{geprags2016origin, cramer2017magnon, ohnuma2013spin}. Furthermore, Mn doping leads to a substantial enhancement of the SSE signal, with the GdIG:Mn (x=0.08) sample exhibiting a significant increase in signal intensity. The pronounced enhancement observed, despite only a modest change in Mn concentration, suggests the involvement of additional underlying mechanisms that merit further investigation.\\

To investigate the origin of the observed SSE signal enhancement, resistivity measurements were performed on all three polycrystalline slabs in the absence of the Pt layer. In each case, the resistance exceeded the instrumental detection limit, confirming that all samples exhibit highly insulating behavior. This effectively rules out the anomalous Nernst effect (ANE) as a contributing factor to the SSE signal\cite{holanda2017longitudinal}. Another factor that may contribute to the enhancement of the SSE is the saturation magnetization $ M_{\text{s}}$. However, in this case, both GdIG and GdIG:Mn samples exhibit similar magnetic behavior with negligible differences. Therefore, the enhancement of the SSE due to changes in saturation magnetization can be ruled out. Previous reports have suggested that doping can influence grain growth, with increased grain size typically enhancing the thermal conductivity ($\kappa$) and, in turn, the SSE response \cite{PhysRevB.95.174401, PhysRevMaterials.1.014601}. Larger grains reduce the density of grain boundaries, thereby minimizing phonon scattering and increasing the phonon mean free path, which results in improved thermal transport. However, in the present study, the crystallite sizes of GdIG and GdIG:Mn samples were found to be comparable, indicating that grain growth is not a significant factor, unlike in systems such as Ce-doped YIG, where notable grain size evolution has been reported \cite{mohmed2019magnetic}.
%The average crystallite size D in sample are calculated from the XRD using the Debye-Scherrer formula, 
%\begin{equation}
%\ D = K {\lambda}/{\beta} cos{\theta}
%  \end{equation}
%where K is the shape factor, \({\lambda}\) wavelength of the X-ray radiation, \({\beta}\) full width at half maximum (FWHM) of the diffraction peak, and \({\theta}\) Bragg angle.
\begin{figure*}
	\includegraphics[width=2.1 \columnwidth]{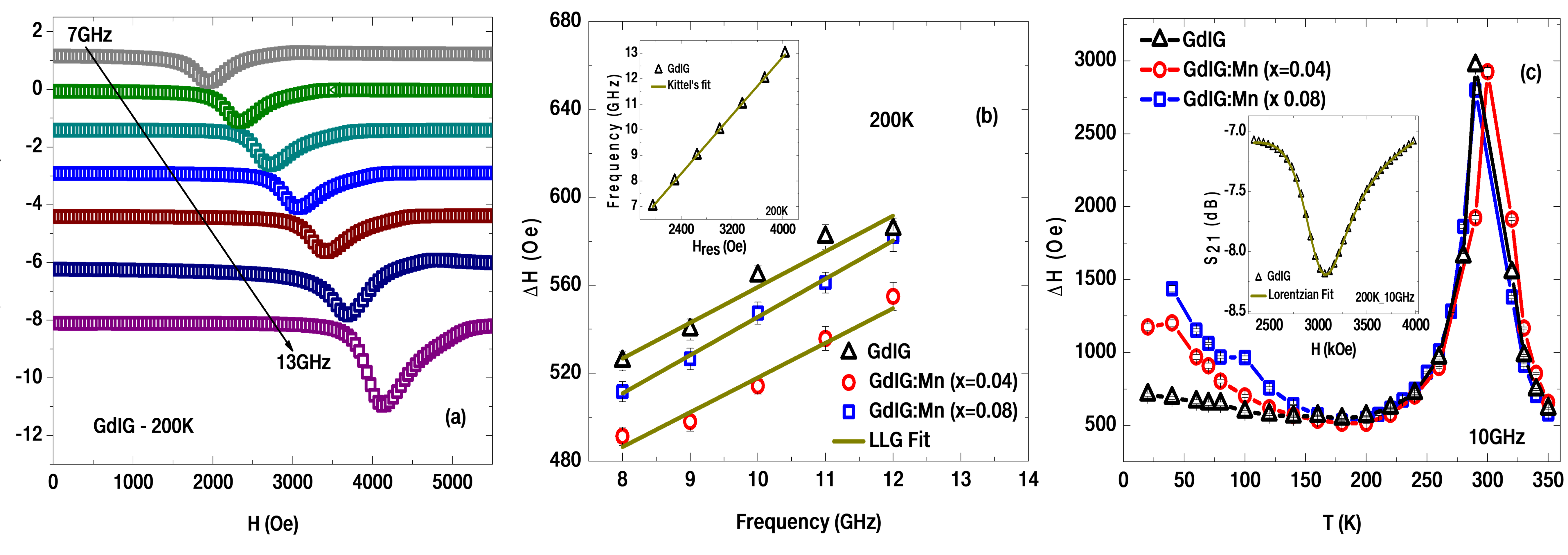}
\centering
    \caption 
{\footnotesize (a) Representative FMR absorption spectra of the GdIG polycrystalline slab measured for various frequencies at 200 K. (b) Frequency-dependent linewidths at 200 K for all three samples, with the inset depicting the Kittel fit for GdIG at 200 K. The effective magnetization and Gilbert damping parameters were extracted using Kittel's equation and the Landau–Lifshitz–Gilbert (LLG) model, respectively, at 200 K. (c) Temperature-dependent ferromagnetic resonance (FMR) linewidths recorded at 10 GHz for GdIG, GdIG:Mn (x=0.04), and GdIG:Mn (x=0.08), with the inset showing a Lorentzian fit for GdIG at 200 K.
}
	\label{FMR} 
\end{figure*}
The average crystallite size D using the Debye-Scherrer formula are $ 5.41 \pm 10^{2}$ nm, $ 4.97 \pm 10^{2}$ nm, $ 5.24\pm 10^{2}$ nm, for GdIG, GdIG:Mn (x=0.04) and GdIG:Mn (x=0.08), respectively. Therefore, any influence of doping-induced grain size changes on phonon-mediated thermal conductivity is unlikely. Additionally, all polycrystalline slabs were prepared under identical sintering conditions, minimizing the likelihood of variations in grain boundaries.\\
\ The quality of the interface is a crucial factor in boosting the SSE signal \cite{ PhysRevLett.107.046601, PhysRevB.87.104412}, influenced by either surface roughness or the presence of a magnetically enriched surface layer. The atomic-scale roughness, interfacial disorder, and chemical mixing can disrupt coherent spin exchange interactions at the interface, reducing the effective spin mixing conductance \({g_{\uparrow\downarrow}}\). Based on first-principles calculations, the spin-mixing conductance  \({g_{\uparrow\downarrow}}\) increases steadily as the magnetic moment density at the normal metal/ferromagnet (NM/FM) interface becomes higher \cite{zhang2011first, pham2018interface}. It has been demonstrated that enhancement of the SSE signal can be achieved through improvements in the interface condition \cite{aqeel2014surface, qiu2015influence, kim2020enhancing, kumawat2024enhanced}. The SSE measurements on YIG with different doping showed an enhancement in signal with higher Fe concentration, and this has been ascribed to an increase in the spin mixing conductance \({g_{\uparrow\downarrow}}\) \cite{PhysRevB.87.104412}. In this study, all polycrystalline slabs were polished using the same technique under identical conditions, and therefore, any variation in signal due to differences in surface quality is unlikely. The possibility of a Mn- and Fe-enriched surface in GdIG:Mn (x = 0.08) compared to GdIG:Mn (x = 0.04) and GdIG cannot be ruled out, suggesting that it can contribute modestly to the observed signal enhancement. However, we believe that this alone does not account for the pronounced increase observed in the total SSE signal.\\
Gilbert damping \({\alpha}\), an intrinsic parameter, plays a crucial role in determining the efficiency of spin current pumping \cite{chang2017role, hoffman2013landau}. We employed temperature-dependent FMR studies on these slabs with the intention of studying the changes in Gilbert damping with respect to Mn doping. The experimental FMR absorption spectra measured for different frequencies at each temperature were fitted using a Lorentzian function to accurately extract the resonance linewidth \({\Delta H}\) and resonance field \(H_{\text{res}}\). The Gilbert damping parameter \({\alpha}\) and inhomogeneous broadening linewidth $\Delta H_{\text{0}}$ are inferred from the slope of the linewidth vs frequency graph using the Landau–Lifshitz–Gilbert equation (LLG) equation as follows,
\begin{equation}
\ {\Delta H(f)}={\Delta H_{\text{0}}} + {4\pi\alpha f}/\gamma
\end{equation}
where ${\Delta H(f)}$ is the frequency-dependent FMR linewidth and $\gamma$ is the gyromagnetic ratio.
To determine the effective magnetization of the sample, the frequency dependence of the resonance field was analyzed using the in-plane Kittel's formula as follows,
\begin{equation}
\ f = \frac{\gamma}{2\pi} \sqrt{H_{\text{res}} (H_{\text{res}} + 4\pi M_{\text{eff}})}
\end{equation}
 Here, $4\pi M_{\text{eff}}$ = $4\pi M_{\text{s}}$ - $ H_{eff}$ is the effective magnetization field and $ H_{eff}$ the effective anistropy field. By fitting the experimental data with the Kittel's equation, the gyromagnetic ratio $\gamma$ and the effective magnetization $\ M_{\text{eff}}$ were extracted. The temperature-dependent FMR studies on polycrystalline slabs of GdIG, GdIG:Mn (x=0.04),  are shown in FIG.\ref{FMR}. The FMR measurements reveal a significant broadening of the linewidth close to the $T_{\text{comp}}$ in all three cases. The FMR linewidth ($\Delta H$) in polycrystalline GdIG slabs can be described as the sum of intrinsic and extrinsic components:  Gilbert damping ($\Delta H_{\text{Gilbert}}$), two-magnon scattering ($\Delta H_{\text{2-magnon}}$) and inhomogeneous broadening ($\Delta H_{\text{0}}$) due to grain boundary scattering, anisotropy variations, and defect-related damping. The total linewidth $\Delta H$ can then be expressed as:
$$
\Delta H = \Delta H_{\text{Gilbert}} + \Delta H_{\text{2-magnon}} + \Delta H_{\text{0}}
$$
Around the $T_{\text{comp}}$, $\Delta H_{\text{2-magnon}}$ and $\Delta H_{\text{0}}$ become the dominant contributions because of reduced magnetization and increased scattering. The relatively narrow linewidth values above and below $T_{\text{comp}}$ can be explained by the dominance of magnon–magnon and magnon–phonon scattering in the magnetic relaxation process. Meanwhile, magnon–electron scattering is minimized due to the lack of conduction electrons in the insulating GdIG slabs \cite{li2022unconventional}.\
\begin{table}%[H] add [H] placement to break table across pages
   \begin{ruledtabular}
 \begin{tabular}{cccccc}
 	\text & \text{\(4\pi M_{\text{eff}}\) (Oe)} & \text{ \(\alpha\) \(\times\) (10$^{-2}$}) & \text{$\Delta H_0$} (Oe)\\
 	\hline
 		GdIG & $1221.6 \pm 5.3$ & $2.28 \pm 0.26 $ & $396.8 \pm 18.3$  \\
 		GdIG:Mn (x=0.04) & $1305.5 \pm 7.7$ & $2.22 \pm 0.27$ & $360.3 \pm 18.4$ \\
 		GdIG:Mn (x=0.08) & $ 1252.1 \pm 5.2$ & $2.42 \pm 0.10$ & $ 371.9 \pm 6.8$ \\
 \end{tabular}
 \end{ruledtabular}
 \caption{\label{FMR Parameters} \footnotesize Comparison of the effective magnetization, Gilbert damping parameters and inhomogeneous broadening linewidth extracted using Kittel's equation and the Landau–Lifshitz–Gilbert (LLG) model at 200 K for different Mn concentration. The variation in the above measured values are minimal regardless of the doping. }
 \end{table}
The inferred values of $4\pi M_{\text{eff}}$ and the Gilbert damping parameter, extracted from fitting at 200K, are presented in TABLE \ref{FMR Parameters}. From the Kittel fit, the obtained $\gamma/2\pi$ value is consistently 2.8MHz/Oe across all three cases. The $M_{\text{eff}}$ and $M_{\text{S}}$ values show minimal variation across GdIG, GdIG:Mn (x=0.04), and GdIG:Mn (x=0.08) samples, indicating that the effective magnetic anisotropy field remains similar among the measured slabs. Based on our calculations, the $H_{\text{eff}}$  values were found to be approximately –465.68 Oe for GdIG, –549.55 Oe for GdIG:Mn (x=0.04) and –496.29 Oe for GdIG:Mn (x=0.08). The negative value signifies easy-plane anisotropy, and this behavior aligns with the expected effect of shape anisotropy in the polycrystalline slab, where the geometry naturally favors in-plane magnetization. Additionally, at 200K, the Gilbert damping parameter for the GdIG sample, determined using the LLG equation, is $2.28 \times 10^{-2}$. The Gilbert damping parameters were observed to be nearly identical in all three cases. Therefore, any contribution from variations in the Gilbert damping to the overall enhancement of the SSE signal is ruled out. We also noted a deviation in the linewidth upon cooling to 20K, with GdIG:Mn (x=0.08) exhibiting a more significant increase than GdIG, as illustrated in FIG.5(c). At low temperatures, when the intrinsic damping decreases, the presence of doping-induced static defects or inhomogeneities can enhance two-magnon scattering, resulting in a broader FMR linewidth\cite{kalarickal2009ferromagnetic,roschmann1977relaxation, roschmann1976two}. Due to the presence of multiple spin-wave branches from various sublattices in GdIG, there is a higher density of degenerate spin-wave modes at a specific energy. The doping-induced defects can more readily connect uniform modes to these complex spin waves, which increases two-magnon scattering, particularly at low temperatures when thermal fluctuations are minimized. Hence, the pronounced change in the FMR linewidth observed in GdIG:Mn (x = 0.08) can be qualitatively related to enhanced two-magnon scattering \cite{nguyen2017two,hurben1998theory,li2024reconfigurable}.\\
\ From the magnon spectra studies of GdIG \cite{geprags2016origin,PhysRevB.101.165137,li2022unconventional}, it has been demonstrated that the acoustic modes ($\alpha$) and gapped optical magnon modes ($\beta$) in the magnon spectrum of GdIG are temperature-dependent and have a direct impact on the efficiency of the SSE. The exchange coupling in GdIG describes the magnetic interactions between spins in the Gd³⁺ and Fe³⁺ sublattices, which are mediated by superexchange via oxygen ions. The strongest interactions take place between octahedral (16a) and tetrahedral (24d) Fe³⁺ ions, characterized by exchange integrals of approximately $J_{\text{ad}} \approx 32 \, \text{cm}^{-1}$ (antiferromagnetic). These dominant interactions determine the Curie temperature ($T_C \approx 560 \, \text{K}$) and establish the ferrimagnetic ordering. The interactions between Gd³⁺ ions and Fe³⁺ ions are weaker, with exchange integrals $J_{\text{ac}} \approx 2\, \text{cm}^{-1}$ and $J_{\text{dc}} \approx 7.00 \, \text{cm}^{-1}$, causing the Gd³⁺ sublattice to be polarized antiparallel to the net Fe³⁺ magnetization. These weaker couplings introduce additional magnon modes and play a role in the compensation temperature\cite{PhysRevB.101.165137, harris1963spin}. The magnon spectrum reflects these exchange interactions, where acoustic modes correspond to the collective magnetization dynamics dominated by Fe–Fe interactions, whereas optical modes arise from inter-sublattice dynamics and are particularly sensitive to Gd–Fe exchange coupling. It has been demonstrated that the substitutional doping of YIG with ions like Ce or Bi \cite{vasili2017direct,imamura2021enhancement} leads to modifications in the Fe–Fe exchange interactions, which in turn significantly affect the SSE response. It is important to recognize the impact of Mn doping on altering the exchange coupling in this context. The strengthened exchange interactions lower the energy of magnon modes or increase their density of states, enabling more magnons to be thermally excited by a temperature gradient, thereby enhancing spin current generation \cite{xie2017first}. Also, at temperatures either above or below the $T_{\text{comp}}$, the stronger exchange coupling enhances magnon contributions from both the low-energy $\alpha$ modes and the high-energy $\beta$ modes, resulting in an increased spin current injected into the Pt layer. This suggests that in Mn-doped GdIG, the exchange coupling between Gd³⁺ and Fe³⁺ is modified due to the replacement of Fe³⁺ ions with Mn³⁺ ions at both octahedral and tetrahedral sites, thereby leading to the observed signal enhancement. While direct measurements of magnon spectra are well established in undoped YIG and GdIG, studies on doped variants remain largely unexplored, hindering a comprehensive understanding of doping-induced modifications in the magnon spectra. This underscores the necessity of magnon spectra studies in doped systems to elucidate the impact of doping on spin dynamics.
\vspace{-0.3em}
\section{Conclusions}
We successfully synthesized polycrystalline samples of GdIG and Mn-doped GdIG via solid-state reaction, with XRD refinement confirming phase purity. Temperature-dependent magnetization measurements revealed a doping-induced shift in compensation temperature, reaching room-temperature compensation for GdIG:Mn (x=0.04), followed by a decrease with higher doping. The temperature-dependent SSE signal approached zero at the magnetic compensation temperature in all three cases, with reversed polarity corresponding to the reversal of the spin quantization axis at the \( T_{\text{comp}}\). Temperature-dependent FMR studies showed negligible variation in Gilbert damping along with similar magnetic characteristics. The enhanced SSE signal strength in Mn-doped samples is attributed to both an increase in spin mixing conductance at the PM/FI interface and due to the modified magnon spectra, likely due to strengthened exchange coupling from Mn doping. As the SSE signal vanishes at the compensation point, layering samples with slightly different compensation temperatures could enable temperature-driven SSE signal rectification. Thin-film heterostructures of these materials hold potential for prototype spin caloritronic devices, such as thermal spin diodes.

\begin{acknowledgments}

\ A.M.F. acknowledges A. De for the scientific discussions. A.M.F. acknowledges DST-INSPIRE for providing financial support through Senior Research Fellowship (SRF). A.M.F. and S.N. acknowledge the funding support by the Ministry of Human Resource Development (MHRD) through the Scheme for Transformational and Advanced Research in Sciences (STARS), and the Department of Science and Technology (DST, Government of India), through grant number CRG/2022/003316.
\\

\end{acknowledgments}

\bibliography{GdIG}

\begin{thebibliography}{53}%
\makeatletter
\providecommand \@ifxundefined [1]{%
 \@ifx{#1\undefined}
}%
\providecommand \@ifnum [1]{%
 \ifnum #1\expandafter \@firstoftwo
 \else \expandafter \@secondoftwo
 \fi
}%
\providecommand \@ifx [1]{%
 \ifx #1\expandafter \@firstoftwo
 \else \expandafter \@secondoftwo
 \fi
}%
\providecommand \natexlab [1]{#1}%
\providecommand \enquote  [1]{``#1''}%
\providecommand \bibnamefont  [1]{#1}%
\providecommand \bibfnamefont [1]{#1}%
\providecommand \citenamefont [1]{#1}%
\providecommand \href@noop [0]{\@secondoftwo}%
\providecommand \href [0]{\begingroup \@sanitize@url \@href}%
\providecommand \@href[1]{\@@startlink{#1}\@@href}%
\providecommand \@@href[1]{\endgroup#1\@@endlink}%
\providecommand \@sanitize@url [0]{\catcode `\\12\catcode `\$12\catcode
  `\&12\catcode `\#12\catcode `\^12\catcode `\_12\catcode `\%12\relax}%
\providecommand \@@startlink[1]{}%
\providecommand \@@endlink[0]{}%
\providecommand \url  [0]{\begingroup\@sanitize@url \@url }%
\providecommand \@url [1]{\endgroup\@href {#1}{\urlprefix }}%
\providecommand \urlprefix  [0]{URL }%
\providecommand \Eprint [0]{\href }%
\providecommand \doibase [0]{https://doi.org/}%
\providecommand \selectlanguage [0]{\@gobble}%
\providecommand \bibinfo  [0]{\@secondoftwo}%
\providecommand \bibfield  [0]{\@secondoftwo}%
\providecommand \translation [1]{[#1]}%
\providecommand \BibitemOpen [0]{}%
\providecommand \bibitemStop [0]{}%
\providecommand \bibitemNoStop [0]{.\EOS\space}%
\providecommand \EOS [0]{\spacefactor3000\relax}%
\providecommand \BibitemShut  [1]{\csname bibitem#1\endcsname}%
\let\auto@bib@innerbib\@empty
%</preamble>
\bibitem [{\citenamefont {Chumak}\ \emph {et~al.}(2015)\citenamefont {Chumak},
  \citenamefont {Vasyuchka}, \citenamefont {Serga},\ and\ \citenamefont
  {Hillebrands}}]{chumak2015magnon}%
  \BibitemOpen
  \bibfield  {author} {\bibinfo {author} {\bibfnamefont {A.~V.}\ \bibnamefont
  {Chumak}}, \bibinfo {author} {\bibfnamefont {V.~I.}\ \bibnamefont
  {Vasyuchka}}, \bibinfo {author} {\bibfnamefont {A.~A.}\ \bibnamefont
  {Serga}},\ and\ \bibinfo {author} {\bibfnamefont {B.}~\bibnamefont
  {Hillebrands}},\ }\href {https://doi.org/https://doi.org/10.1038/nphys3347}
  {\bibfield  {journal} {\bibinfo  {journal} {Nature Physics}\ }\textbf
  {\bibinfo {volume} {11}},\ \bibinfo {pages} {453} (\bibinfo {year}
  {2015})}\BibitemShut {NoStop}%
\bibitem [{\citenamefont {Pirro}\ \emph {et~al.}(2021)\citenamefont {Pirro},
  \citenamefont {Vasyuchka}, \citenamefont {Serga},\ and\ \citenamefont
  {Hillebrands}}]{pirro2021advances}%
  \BibitemOpen
  \bibfield  {author} {\bibinfo {author} {\bibfnamefont {P.}~\bibnamefont
  {Pirro}}, \bibinfo {author} {\bibfnamefont {V.~I.}\ \bibnamefont
  {Vasyuchka}}, \bibinfo {author} {\bibfnamefont {A.~A.}\ \bibnamefont
  {Serga}},\ and\ \bibinfo {author} {\bibfnamefont {B.}~\bibnamefont
  {Hillebrands}},\ }\href
  {https://doi.org/https://doi.org/10.1038/s41578-021-00332-w} {\bibfield
  {journal} {\bibinfo  {journal} {Nature Reviews Materials}\ }\textbf {\bibinfo
  {volume} {6}},\ \bibinfo {pages} {1114} (\bibinfo {year} {2021})}\BibitemShut
  {NoStop}%
\bibitem [{\citenamefont {Kehlberger}\ \emph {et~al.}(2015)\citenamefont
  {Kehlberger}, \citenamefont {Ritzmann}, \citenamefont {Hinzke}, \citenamefont
  {Guo}, \citenamefont {Cramer}, \citenamefont {Jakob}, \citenamefont
  {Onbasli}, \citenamefont {Kim}, \citenamefont {Ross}, \citenamefont
  {Jungfleisch}, \citenamefont {Hillebrands}, \citenamefont {Nowak},\ and\
  \citenamefont {Kl\"aui}}]{PhysRevLett.115.096602}%
  \BibitemOpen
  \bibfield  {author} {\bibinfo {author} {\bibfnamefont {A.}~\bibnamefont
  {Kehlberger}}, \bibinfo {author} {\bibfnamefont {U.}~\bibnamefont
  {Ritzmann}}, \bibinfo {author} {\bibfnamefont {D.}~\bibnamefont {Hinzke}},
  \bibinfo {author} {\bibfnamefont {E.-J.}\ \bibnamefont {Guo}}, \bibinfo
  {author} {\bibfnamefont {J.}~\bibnamefont {Cramer}}, \bibinfo {author}
  {\bibfnamefont {G.}~\bibnamefont {Jakob}}, \bibinfo {author} {\bibfnamefont
  {M.~C.}\ \bibnamefont {Onbasli}}, \bibinfo {author} {\bibfnamefont {D.~H.}\
  \bibnamefont {Kim}}, \bibinfo {author} {\bibfnamefont {C.~A.}\ \bibnamefont
  {Ross}}, \bibinfo {author} {\bibfnamefont {M.~B.}\ \bibnamefont
  {Jungfleisch}}, \bibinfo {author} {\bibfnamefont {B.}~\bibnamefont
  {Hillebrands}}, \bibinfo {author} {\bibfnamefont {U.}~\bibnamefont {Nowak}},\
  and\ \bibinfo {author} {\bibfnamefont {M.}~\bibnamefont {Kl\"aui}},\ }\href
  {https://doi.org/10.1103/PhysRevLett.115.096602} {\bibfield  {journal}
  {\bibinfo  {journal} {Phys. Rev. Lett.}\ }\textbf {\bibinfo {volume} {115}},\
  \bibinfo {pages} {096602} (\bibinfo {year} {2015})}\BibitemShut {NoStop}%
\bibitem [{\citenamefont {Serga}\ \emph {et~al.}(2010)\citenamefont {Serga},
  \citenamefont {Chumak},\ and\ \citenamefont {Hillebrands}}]{serga2010yig}%
  \BibitemOpen
  \bibfield  {author} {\bibinfo {author} {\bibfnamefont {A.}~\bibnamefont
  {Serga}}, \bibinfo {author} {\bibfnamefont {A.}~\bibnamefont {Chumak}},\ and\
  \bibinfo {author} {\bibfnamefont {B.}~\bibnamefont {Hillebrands}},\ }\href
  {https://doi.org/10.1088/0022-3727/43/26/264002} {\bibfield  {journal}
  {\bibinfo  {journal} {Journal of Physics D: Applied Physics}\ }\textbf
  {\bibinfo {volume} {43}},\ \bibinfo {pages} {264002} (\bibinfo {year}
  {2010})}\BibitemShut {NoStop}%
\bibitem [{\citenamefont {Uchida}\ \emph {et~al.}(2008)\citenamefont {Uchida},
  \citenamefont {Takahashi}, \citenamefont {Harii}, \citenamefont {Ieda},
  \citenamefont {Koshibae}, \citenamefont {Ando}, \citenamefont {Maekawa},\
  and\ \citenamefont {Saitoh}}]{uchida2008observation}%
  \BibitemOpen
  \bibfield  {author} {\bibinfo {author} {\bibfnamefont {K.-I.}\ \bibnamefont
  {Uchida}}, \bibinfo {author} {\bibfnamefont {S.}~\bibnamefont {Takahashi}},
  \bibinfo {author} {\bibfnamefont {K.}~\bibnamefont {Harii}}, \bibinfo
  {author} {\bibfnamefont {J.}~\bibnamefont {Ieda}}, \bibinfo {author}
  {\bibfnamefont {W.}~\bibnamefont {Koshibae}}, \bibinfo {author}
  {\bibfnamefont {K.}~\bibnamefont {Ando}}, \bibinfo {author} {\bibfnamefont
  {S.}~\bibnamefont {Maekawa}},\ and\ \bibinfo {author} {\bibfnamefont
  {E.}~\bibnamefont {Saitoh}},\ }\href
  {https://doi.org/https://doi.org/10.1038/nature07321} {\bibfield  {journal}
  {\bibinfo  {journal} {Nature}\ }\textbf {\bibinfo {volume} {455}},\ \bibinfo
  {pages} {778} (\bibinfo {year} {2008})}\BibitemShut {NoStop}%
\bibitem [{\citenamefont {Bauer}\ \emph {et~al.}(2012)\citenamefont {Bauer},
  \citenamefont {Saitoh},\ and\ \citenamefont {Van~Wees}}]{bauer2012spin}%
  \BibitemOpen
  \bibfield  {author} {\bibinfo {author} {\bibfnamefont {G.~E.}\ \bibnamefont
  {Bauer}}, \bibinfo {author} {\bibfnamefont {E.}~\bibnamefont {Saitoh}},\ and\
  \bibinfo {author} {\bibfnamefont {B.~J.}\ \bibnamefont {Van~Wees}},\ }\href
  {https://doi.org/https://doi.org/10.1038/nmat3301} {\bibfield  {journal}
  {\bibinfo  {journal} {Nature Materials}\ }\textbf {\bibinfo {volume} {11}},\
  \bibinfo {pages} {391} (\bibinfo {year} {2012})}\BibitemShut {NoStop}%
\bibitem [{\citenamefont {Jaworski}\ \emph {et~al.}(2010)\citenamefont
  {Jaworski}, \citenamefont {Yang}, \citenamefont {Mack}, \citenamefont
  {Awschalom}, \citenamefont {Heremans},\ and\ \citenamefont
  {Myers}}]{jaworski2010observation}%
  \BibitemOpen
  \bibfield  {author} {\bibinfo {author} {\bibfnamefont {C.}~\bibnamefont
  {Jaworski}}, \bibinfo {author} {\bibfnamefont {J.}~\bibnamefont {Yang}},
  \bibinfo {author} {\bibfnamefont {S.}~\bibnamefont {Mack}}, \bibinfo {author}
  {\bibfnamefont {D.}~\bibnamefont {Awschalom}}, \bibinfo {author}
  {\bibfnamefont {J.}~\bibnamefont {Heremans}},\ and\ \bibinfo {author}
  {\bibfnamefont {R.}~\bibnamefont {Myers}},\ }\href
  {https://doi.org/https://doi.org/10.1038/nmat2860} {\bibfield  {journal}
  {\bibinfo  {journal} {Nature Materials}\ }\textbf {\bibinfo {volume} {9}},\
  \bibinfo {pages} {898} (\bibinfo {year} {2010})}\BibitemShut {NoStop}%
\bibitem [{\citenamefont {Uchida}\ \emph
  {et~al.}(2010{\natexlab{a}})\citenamefont {Uchida}, \citenamefont {Xiao},
  \citenamefont {Adachi}, \citenamefont {Ohe}, \citenamefont {Takahashi},
  \citenamefont {Ieda}, \citenamefont {Ota}, \citenamefont {Kajiwara},
  \citenamefont {Umezawa}, \citenamefont {Kawai} \emph
  {et~al.}}]{uchida2010spin}%
  \BibitemOpen
  \bibfield  {author} {\bibinfo {author} {\bibfnamefont {K.-i.}\ \bibnamefont
  {Uchida}}, \bibinfo {author} {\bibfnamefont {J.}~\bibnamefont {Xiao}},
  \bibinfo {author} {\bibfnamefont {H.}~\bibnamefont {Adachi}}, \bibinfo
  {author} {\bibfnamefont {J.-i.}\ \bibnamefont {Ohe}}, \bibinfo {author}
  {\bibfnamefont {S.}~\bibnamefont {Takahashi}}, \bibinfo {author}
  {\bibfnamefont {J.}~\bibnamefont {Ieda}}, \bibinfo {author} {\bibfnamefont
  {T.}~\bibnamefont {Ota}}, \bibinfo {author} {\bibfnamefont {Y.}~\bibnamefont
  {Kajiwara}}, \bibinfo {author} {\bibfnamefont {H.}~\bibnamefont {Umezawa}},
  \bibinfo {author} {\bibfnamefont {H.}~\bibnamefont {Kawai}}, \emph {et~al.},\
  }\href {https://doi.org/https://doi.org/10.1038/nmat2856} {\bibfield
  {journal} {\bibinfo  {journal} {Nature Materials}\ }\textbf {\bibinfo
  {volume} {9}},\ \bibinfo {pages} {894} (\bibinfo {year}
  {2010}{\natexlab{a}})}\BibitemShut {NoStop}%
\bibitem [{\citenamefont {Adachi}\ \emph {et~al.}(2013)\citenamefont {Adachi},
  \citenamefont {Uchida}, \citenamefont {Saitoh},\ and\ \citenamefont
  {Maekawa}}]{adachi2013theory}%
  \BibitemOpen
  \bibfield  {author} {\bibinfo {author} {\bibfnamefont {H.}~\bibnamefont
  {Adachi}}, \bibinfo {author} {\bibfnamefont {K.-i.}\ \bibnamefont {Uchida}},
  \bibinfo {author} {\bibfnamefont {E.}~\bibnamefont {Saitoh}},\ and\ \bibinfo
  {author} {\bibfnamefont {S.}~\bibnamefont {Maekawa}},\ }\href
  {https://doi.org/10.1088/0034-4885/76/3/036501} {\bibfield  {journal}
  {\bibinfo  {journal} {Reports on Progress in Physics}\ }\textbf {\bibinfo
  {volume} {76}},\ \bibinfo {pages} {036501} (\bibinfo {year}
  {2013})}\BibitemShut {NoStop}%
\bibitem [{\citenamefont {Wu}\ \emph {et~al.}(2016)\citenamefont {Wu},
  \citenamefont {Zhang}, \citenamefont {Kc}, \citenamefont {Borisov},
  \citenamefont {Pearson}, \citenamefont {Jiang}, \citenamefont {Lederman},
  \citenamefont {Hoffmann},\ and\ \citenamefont
  {Bhattacharya}}]{wu2016antiferromagnetic}%
  \BibitemOpen
  \bibfield  {author} {\bibinfo {author} {\bibfnamefont {S.~M.}\ \bibnamefont
  {Wu}}, \bibinfo {author} {\bibfnamefont {W.}~\bibnamefont {Zhang}}, \bibinfo
  {author} {\bibfnamefont {A.}~\bibnamefont {Kc}}, \bibinfo {author}
  {\bibfnamefont {P.}~\bibnamefont {Borisov}}, \bibinfo {author} {\bibfnamefont
  {J.~E.}\ \bibnamefont {Pearson}}, \bibinfo {author} {\bibfnamefont {J.~S.}\
  \bibnamefont {Jiang}}, \bibinfo {author} {\bibfnamefont {D.}~\bibnamefont
  {Lederman}}, \bibinfo {author} {\bibfnamefont {A.}~\bibnamefont {Hoffmann}},\
  and\ \bibinfo {author} {\bibfnamefont {A.}~\bibnamefont {Bhattacharya}},\
  }\href {https://doi.org/https://doi.org/10.1103/PhysRevLett.116.097204}
  {\bibfield  {journal} {\bibinfo  {journal} {Physical Review Letters}\
  }\textbf {\bibinfo {volume} {116}},\ \bibinfo {pages} {097204} (\bibinfo
  {year} {2016})}\BibitemShut {NoStop}%
\bibitem [{\citenamefont {Rezende}\ \emph {et~al.}(2016)\citenamefont
  {Rezende}, \citenamefont {Rodr{\'\i}guez-Su{\'a}rez},\ and\ \citenamefont
  {Azevedo}}]{rezende2016theory}%
  \BibitemOpen
  \bibfield  {author} {\bibinfo {author} {\bibfnamefont {S.}~\bibnamefont
  {Rezende}}, \bibinfo {author} {\bibfnamefont {R.}~\bibnamefont
  {Rodr{\'\i}guez-Su{\'a}rez}},\ and\ \bibinfo {author} {\bibfnamefont
  {A.}~\bibnamefont {Azevedo}},\ }\href
  {https://doi.org/https://doi.org/10.1103/PhysRevB.93.014425} {\bibfield
  {journal} {\bibinfo  {journal} {Physical Review B}\ }\textbf {\bibinfo
  {volume} {93}},\ \bibinfo {pages} {014425} (\bibinfo {year}
  {2016})}\BibitemShut {NoStop}%
\bibitem [{\citenamefont {Ohnuma}\ \emph
  {et~al.}(2013{\natexlab{a}})\citenamefont {Ohnuma}, \citenamefont {Adachi},
  \citenamefont {Saitoh},\ and\ \citenamefont {Maekawa}}]{PhysRevB.87.014423}%
  \BibitemOpen
  \bibfield  {author} {\bibinfo {author} {\bibfnamefont {Y.}~\bibnamefont
  {Ohnuma}}, \bibinfo {author} {\bibfnamefont {H.}~\bibnamefont {Adachi}},
  \bibinfo {author} {\bibfnamefont {E.}~\bibnamefont {Saitoh}},\ and\ \bibinfo
  {author} {\bibfnamefont {S.}~\bibnamefont {Maekawa}},\ }\href
  {https://doi.org/10.1103/PhysRevB.87.014423} {\bibfield  {journal} {\bibinfo
  {journal} {Phys. Rev. B}\ }\textbf {\bibinfo {volume} {87}},\ \bibinfo
  {pages} {014423} (\bibinfo {year} {2013}{\natexlab{a}})}\BibitemShut
  {NoStop}%
\bibitem [{\citenamefont {Jungwirth}\ \emph {et~al.}(2016)\citenamefont
  {Jungwirth}, \citenamefont {Marti}, \citenamefont {Wadley},\ and\
  \citenamefont {Wunderlich}}]{jungwirth2016antiferromagnetic}%
  \BibitemOpen
  \bibfield  {author} {\bibinfo {author} {\bibfnamefont {T.}~\bibnamefont
  {Jungwirth}}, \bibinfo {author} {\bibfnamefont {X.}~\bibnamefont {Marti}},
  \bibinfo {author} {\bibfnamefont {P.}~\bibnamefont {Wadley}},\ and\ \bibinfo
  {author} {\bibfnamefont {J.}~\bibnamefont {Wunderlich}},\ }\href
  {https://doi.org/10.1038/NNANO.2016.18} {\bibfield  {journal} {\bibinfo
  {journal} {Nature nanotechnology}\ }\textbf {\bibinfo {volume} {11}},\
  \bibinfo {pages} {231} (\bibinfo {year} {2016})}\BibitemShut {NoStop}%
\bibitem [{\citenamefont {Puebla}\ \emph {et~al.}(2020)\citenamefont {Puebla},
  \citenamefont {Kim}, \citenamefont {Kondou},\ and\ \citenamefont
  {Otani}}]{puebla2020spintronic}%
  \BibitemOpen
  \bibfield  {author} {\bibinfo {author} {\bibfnamefont {J.}~\bibnamefont
  {Puebla}}, \bibinfo {author} {\bibfnamefont {J.}~\bibnamefont {Kim}},
  \bibinfo {author} {\bibfnamefont {K.}~\bibnamefont {Kondou}},\ and\ \bibinfo
  {author} {\bibfnamefont {Y.}~\bibnamefont {Otani}},\ }\href
  {https://doi.org/https://doi.org/10.1038/s43246-020-0022-5} {\bibfield
  {journal} {\bibinfo  {journal} {Communications Materials}\ }\textbf {\bibinfo
  {volume} {1}},\ \bibinfo {pages} {24} (\bibinfo {year} {2020})}\BibitemShut
  {NoStop}%
\bibitem [{\citenamefont {Kim}\ \emph {et~al.}(2022)\citenamefont {Kim},
  \citenamefont {Beach}, \citenamefont {Lee}, \citenamefont {Ono},
  \citenamefont {Rasing},\ and\ \citenamefont {Yang}}]{kim2022ferrimagnetic}%
  \BibitemOpen
  \bibfield  {author} {\bibinfo {author} {\bibfnamefont {S.~K.}\ \bibnamefont
  {Kim}}, \bibinfo {author} {\bibfnamefont {G.~S.}\ \bibnamefont {Beach}},
  \bibinfo {author} {\bibfnamefont {K.-J.}\ \bibnamefont {Lee}}, \bibinfo
  {author} {\bibfnamefont {T.}~\bibnamefont {Ono}}, \bibinfo {author}
  {\bibfnamefont {T.}~\bibnamefont {Rasing}},\ and\ \bibinfo {author}
  {\bibfnamefont {H.}~\bibnamefont {Yang}},\ }\href
  {https://doi.org/https://doi.org/10.1038/s41563-021-01139-4} {\bibfield
  {journal} {\bibinfo  {journal} {Nature materials}\ }\textbf {\bibinfo
  {volume} {21}},\ \bibinfo {pages} {24} (\bibinfo {year} {2022})}\BibitemShut
  {NoStop}%
\bibitem [{\citenamefont {Gepr{\"a}gs}\ \emph {et~al.}(2016)\citenamefont
  {Gepr{\"a}gs}, \citenamefont {Kehlberger}, \citenamefont {Coletta},
  \citenamefont {Qiu}, \citenamefont {Guo}, \citenamefont {Schulz},
  \citenamefont {Mix}, \citenamefont {Meyer}, \citenamefont {Kamra},
  \citenamefont {Althammer} \emph {et~al.}}]{geprags2016origin}%
  \BibitemOpen
  \bibfield  {author} {\bibinfo {author} {\bibfnamefont {S.}~\bibnamefont
  {Gepr{\"a}gs}}, \bibinfo {author} {\bibfnamefont {A.}~\bibnamefont
  {Kehlberger}}, \bibinfo {author} {\bibfnamefont {F.~D.}\ \bibnamefont
  {Coletta}}, \bibinfo {author} {\bibfnamefont {Z.}~\bibnamefont {Qiu}},
  \bibinfo {author} {\bibfnamefont {E.-J.}\ \bibnamefont {Guo}}, \bibinfo
  {author} {\bibfnamefont {T.}~\bibnamefont {Schulz}}, \bibinfo {author}
  {\bibfnamefont {C.}~\bibnamefont {Mix}}, \bibinfo {author} {\bibfnamefont
  {S.}~\bibnamefont {Meyer}}, \bibinfo {author} {\bibfnamefont
  {A.}~\bibnamefont {Kamra}}, \bibinfo {author} {\bibfnamefont
  {M.}~\bibnamefont {Althammer}}, \emph {et~al.},\ }\href
  {https://doi.org/https://doi.org/10.1038/ncomms10452} {\bibfield  {journal}
  {\bibinfo  {journal} {Nature Communications}\ }\textbf {\bibinfo {volume}
  {7}},\ \bibinfo {pages} {10452} (\bibinfo {year} {2016})}\BibitemShut
  {NoStop}%
\bibitem [{\citenamefont {Holzmann}\ \emph {et~al.}(2022)\citenamefont
  {Holzmann}, \citenamefont {Ullrich}, \citenamefont {Ciubotariu},\ and\
  \citenamefont {Albrecht}}]{holzmann2022stress}%
  \BibitemOpen
  \bibfield  {author} {\bibinfo {author} {\bibfnamefont {C.}~\bibnamefont
  {Holzmann}}, \bibinfo {author} {\bibfnamefont {A.}~\bibnamefont {Ullrich}},
  \bibinfo {author} {\bibfnamefont {O.-T.}\ \bibnamefont {Ciubotariu}},\ and\
  \bibinfo {author} {\bibfnamefont {M.}~\bibnamefont {Albrecht}},\ }\href
  {https://doi.org/https://doi.org/10.1021/acsanm.1c03687} {\bibfield
  {journal} {\bibinfo  {journal} {ACS Applied Nano Materials}\ }\textbf
  {\bibinfo {volume} {5}},\ \bibinfo {pages} {1023} (\bibinfo {year}
  {2022})}\BibitemShut {NoStop}%
\bibitem [{\citenamefont {Liang}\ \emph {et~al.}(2023)\citenamefont {Liang},
  \citenamefont {Zhao}, \citenamefont {Liu}, \citenamefont {Li}, \citenamefont
  {Ng}, \citenamefont {Wong}, \citenamefont {Cheng}, \citenamefont {Zhou},
  \citenamefont {Dai}, \citenamefont {Mak} \emph
  {et~al.}}]{liang2023thickness}%
  \BibitemOpen
  \bibfield  {author} {\bibinfo {author} {\bibfnamefont {J.~M.}\ \bibnamefont
  {Liang}}, \bibinfo {author} {\bibfnamefont {X.~W.}\ \bibnamefont {Zhao}},
  \bibinfo {author} {\bibfnamefont {Y.~K.}\ \bibnamefont {Liu}}, \bibinfo
  {author} {\bibfnamefont {P.~G.}\ \bibnamefont {Li}}, \bibinfo {author}
  {\bibfnamefont {S.~M.}\ \bibnamefont {Ng}}, \bibinfo {author} {\bibfnamefont
  {H.~F.}\ \bibnamefont {Wong}}, \bibinfo {author} {\bibfnamefont {W.~F.}\
  \bibnamefont {Cheng}}, \bibinfo {author} {\bibfnamefont {Y.}~\bibnamefont
  {Zhou}}, \bibinfo {author} {\bibfnamefont {J.~Y.}\ \bibnamefont {Dai}},
  \bibinfo {author} {\bibfnamefont {C.~L.}\ \bibnamefont {Mak}}, \emph
  {et~al.},\ }\href
  {https://pubs.aip.org/aip/apl/article/122/24/242401/2895963/The-thickness-effect-on-the-compensation}
  {\bibfield  {journal} {\bibinfo  {journal} {Applied Physics Letters}\
  }\textbf {\bibinfo {volume} {122}} (\bibinfo {year} {2023})}\BibitemShut
  {NoStop}%
\bibitem [{\citenamefont {Chanda}\ \emph {et~al.}(2022)\citenamefont {Chanda},
  \citenamefont {Holzmann}, \citenamefont {Schulz}, \citenamefont {Seyd},
  \citenamefont {Albrecht}, \citenamefont {Phan},\ and\ \citenamefont
  {Srikanth}}]{chanda2022scaling}%
  \BibitemOpen
  \bibfield  {author} {\bibinfo {author} {\bibfnamefont {A.}~\bibnamefont
  {Chanda}}, \bibinfo {author} {\bibfnamefont {C.}~\bibnamefont {Holzmann}},
  \bibinfo {author} {\bibfnamefont {N.}~\bibnamefont {Schulz}}, \bibinfo
  {author} {\bibfnamefont {J.}~\bibnamefont {Seyd}}, \bibinfo {author}
  {\bibfnamefont {M.}~\bibnamefont {Albrecht}}, \bibinfo {author}
  {\bibfnamefont {M.-H.}\ \bibnamefont {Phan}},\ and\ \bibinfo {author}
  {\bibfnamefont {H.}~\bibnamefont {Srikanth}},\ }\href
  {https://doi.org/https://doi.org/10.1002/adfm.202109170} {\bibfield
  {journal} {\bibinfo  {journal} {Advanced Functional Materials}\ }\textbf
  {\bibinfo {volume} {32}},\ \bibinfo {pages} {2109170} (\bibinfo {year}
  {2022})}\BibitemShut {NoStop}%
\bibitem [{\citenamefont {Chanda}\ \emph {et~al.}(2024)\citenamefont {Chanda},
  \citenamefont {Holzmann}, \citenamefont {Schulz}, \citenamefont {Stein},
  \citenamefont {Albrecht}, \citenamefont {Phan},\ and\ \citenamefont
  {Srikanth}}]{chanda2024thermally}%
  \BibitemOpen
  \bibfield  {author} {\bibinfo {author} {\bibfnamefont {A.}~\bibnamefont
  {Chanda}}, \bibinfo {author} {\bibfnamefont {C.}~\bibnamefont {Holzmann}},
  \bibinfo {author} {\bibfnamefont {N.}~\bibnamefont {Schulz}}, \bibinfo
  {author} {\bibfnamefont {D.}~\bibnamefont {Stein}}, \bibinfo {author}
  {\bibfnamefont {M.}~\bibnamefont {Albrecht}}, \bibinfo {author}
  {\bibfnamefont {M.-H.}\ \bibnamefont {Phan}},\ and\ \bibinfo {author}
  {\bibfnamefont {H.}~\bibnamefont {Srikanth}},\ }\href
  {https://pubs.aip.org/aip/apl/article/122/24/242401/2895963/The-thickness-effect-on-the-compensation}
  {\bibfield  {journal} {\bibinfo  {journal} {Journal of Applied Physics}\
  }\textbf {\bibinfo {volume} {135}} (\bibinfo {year} {2024})}\BibitemShut
  {NoStop}%
\bibitem [{\citenamefont {Wang}\ \emph {et~al.}(2017)\citenamefont {Wang},
  \citenamefont {Liang}, \citenamefont {Zhang}, \citenamefont {Liang},
  \citenamefont {Zhu}, \citenamefont {Qin}, \citenamefont {Gao}, \citenamefont
  {Peng}, \citenamefont {Sun},\ and\ \citenamefont {Bi}}]{PhysRevB.96.224403}%
  \BibitemOpen
  \bibfield  {author} {\bibinfo {author} {\bibfnamefont {C.~T.}\ \bibnamefont
  {Wang}}, \bibinfo {author} {\bibfnamefont {X.~F.}\ \bibnamefont {Liang}},
  \bibinfo {author} {\bibfnamefont {Y.}~\bibnamefont {Zhang}}, \bibinfo
  {author} {\bibfnamefont {X.}~\bibnamefont {Liang}}, \bibinfo {author}
  {\bibfnamefont {Y.~P.}\ \bibnamefont {Zhu}}, \bibinfo {author} {\bibfnamefont
  {J.}~\bibnamefont {Qin}}, \bibinfo {author} {\bibfnamefont {Y.}~\bibnamefont
  {Gao}}, \bibinfo {author} {\bibfnamefont {B.}~\bibnamefont {Peng}}, \bibinfo
  {author} {\bibfnamefont {N.~X.}\ \bibnamefont {Sun}},\ and\ \bibinfo {author}
  {\bibfnamefont {L.}~\bibnamefont {Bi}},\ }\href
  {https://doi.org/10.1103/PhysRevB.96.224403} {\bibfield  {journal} {\bibinfo
  {journal} {Phys. Rev. B}\ }\textbf {\bibinfo {volume} {96}},\ \bibinfo
  {pages} {224403} (\bibinfo {year} {2017})}\BibitemShut {NoStop}%
\bibitem [{\citenamefont {Wehmeyer}\ \emph {et~al.}(2017)\citenamefont
  {Wehmeyer}, \citenamefont {Yabuki}, \citenamefont {Monachon}, \citenamefont
  {Wu},\ and\ \citenamefont {Dames}}]{wehmeyer2017thermal}%
  \BibitemOpen
  \bibfield  {author} {\bibinfo {author} {\bibfnamefont {G.}~\bibnamefont
  {Wehmeyer}}, \bibinfo {author} {\bibfnamefont {T.}~\bibnamefont {Yabuki}},
  \bibinfo {author} {\bibfnamefont {C.}~\bibnamefont {Monachon}}, \bibinfo
  {author} {\bibfnamefont {J.}~\bibnamefont {Wu}},\ and\ \bibinfo {author}
  {\bibfnamefont {C.}~\bibnamefont {Dames}},\ }\href
  {https://pubs.aip.org/aip/apr/article/4/4/041304/123852/Thermal-diodes-regulators-and-switches-Physical}
  {\bibfield  {journal} {\bibinfo  {journal} {Applied Physics Reviews}\
  }\textbf {\bibinfo {volume} {4}} (\bibinfo {year} {2017})}\BibitemShut
  {NoStop}%
\bibitem [{\citenamefont {Finocchio}\ \emph {et~al.}(2021)\citenamefont
  {Finocchio}, \citenamefont {Tomasello}, \citenamefont {Fang}, \citenamefont
  {Giordano}, \citenamefont {Puliafito}, \citenamefont {Carpentieri},\ and\
  \citenamefont {Zeng}}]{finocchio2021perspectives}%
  \BibitemOpen
  \bibfield  {author} {\bibinfo {author} {\bibfnamefont {G.}~\bibnamefont
  {Finocchio}}, \bibinfo {author} {\bibfnamefont {R.}~\bibnamefont
  {Tomasello}}, \bibinfo {author} {\bibfnamefont {B.}~\bibnamefont {Fang}},
  \bibinfo {author} {\bibfnamefont {A.}~\bibnamefont {Giordano}}, \bibinfo
  {author} {\bibfnamefont {V.}~\bibnamefont {Puliafito}}, \bibinfo {author}
  {\bibfnamefont {M.}~\bibnamefont {Carpentieri}},\ and\ \bibinfo {author}
  {\bibfnamefont {Z.}~\bibnamefont {Zeng}},\ }\href
  {https://pubs.aip.org/aip/apl/article/118/16/160502/1062192/Perspectives-on-spintronic-diodes}
  {\bibfield  {journal} {\bibinfo  {journal} {Applied Physics Letters}\
  }\textbf {\bibinfo {volume} {118}} (\bibinfo {year} {2021})}\BibitemShut
  {NoStop}%
\bibitem [{\citenamefont {De}\ \emph {et~al.}(2020)\citenamefont {De},
  \citenamefont {Ghosh}, \citenamefont {Mandal}, \citenamefont {Ogale},\ and\
  \citenamefont {Nair}}]{PhysRevLett.124.017203}%
  \BibitemOpen
  \bibfield  {author} {\bibinfo {author} {\bibfnamefont {A.}~\bibnamefont
  {De}}, \bibinfo {author} {\bibfnamefont {A.}~\bibnamefont {Ghosh}}, \bibinfo
  {author} {\bibfnamefont {R.}~\bibnamefont {Mandal}}, \bibinfo {author}
  {\bibfnamefont {S.}~\bibnamefont {Ogale}},\ and\ \bibinfo {author}
  {\bibfnamefont {S.}~\bibnamefont {Nair}},\ }\href
  {https://doi.org/10.1103/PhysRevLett.124.017203} {\bibfield  {journal}
  {\bibinfo  {journal} {Phys. Rev. Lett.}\ }\textbf {\bibinfo {volume} {124}},\
  \bibinfo {pages} {017203} (\bibinfo {year} {2020})}\BibitemShut {NoStop}%
\bibitem [{\citenamefont {Uchida}\ \emph
  {et~al.}(2010{\natexlab{b}})\citenamefont {Uchida}, \citenamefont {Adachi},
  \citenamefont {Ota}, \citenamefont {Nakayama}, \citenamefont {Maekawa},\ and\
  \citenamefont {Saitoh}}]{uchida2010observation}%
  \BibitemOpen
  \bibfield  {author} {\bibinfo {author} {\bibfnamefont {K.-i.}\ \bibnamefont
  {Uchida}}, \bibinfo {author} {\bibfnamefont {H.}~\bibnamefont {Adachi}},
  \bibinfo {author} {\bibfnamefont {T.}~\bibnamefont {Ota}}, \bibinfo {author}
  {\bibfnamefont {H.}~\bibnamefont {Nakayama}}, \bibinfo {author}
  {\bibfnamefont {S.}~\bibnamefont {Maekawa}},\ and\ \bibinfo {author}
  {\bibfnamefont {E.}~\bibnamefont {Saitoh}},\ }\href
  {https://pubs.aip.org/aip/apl/article/97/17/172505/121243/Observation-of-longitudinal-spin-Seebeck-effect-in}
  {\bibfield  {journal} {\bibinfo  {journal} {Applied Physics Letters}\
  }\textbf {\bibinfo {volume} {97}} (\bibinfo {year}
  {2010}{\natexlab{b}})}\BibitemShut {NoStop}%
\bibitem [{\citenamefont {Cramer}\ \emph {et~al.}(2017)\citenamefont {Cramer},
  \citenamefont {Guo}, \citenamefont {Geprägs}, \citenamefont {Kehlberger},
  \citenamefont {Ivanov}, \citenamefont {Ganzhorn}, \citenamefont
  {Della~Coletta}, \citenamefont {Althammer}, \citenamefont {Huebl},
  \citenamefont {Gross} \emph {et~al.}}]{cramer2017magnon}%
  \BibitemOpen
  \bibfield  {author} {\bibinfo {author} {\bibfnamefont {J.}~\bibnamefont
  {Cramer}}, \bibinfo {author} {\bibfnamefont {E.-J.}\ \bibnamefont {Guo}},
  \bibinfo {author} {\bibfnamefont {S.}~\bibnamefont {Geprägs}}, \bibinfo
  {author} {\bibfnamefont {A.}~\bibnamefont {Kehlberger}}, \bibinfo {author}
  {\bibfnamefont {Y.~P.}\ \bibnamefont {Ivanov}}, \bibinfo {author}
  {\bibfnamefont {K.}~\bibnamefont {Ganzhorn}}, \bibinfo {author}
  {\bibfnamefont {F.}~\bibnamefont {Della~Coletta}}, \bibinfo {author}
  {\bibfnamefont {M.}~\bibnamefont {Althammer}}, \bibinfo {author}
  {\bibfnamefont {H.}~\bibnamefont {Huebl}}, \bibinfo {author} {\bibfnamefont
  {R.}~\bibnamefont {Gross}}, \emph {et~al.},\ }\href
  {https://doi.org/https://doi.org/10.1021/acs.nanolett.6b04522} {\bibfield
  {journal} {\bibinfo  {journal} {Nano Letters}\ }\textbf {\bibinfo {volume}
  {17}},\ \bibinfo {pages} {3334} (\bibinfo {year} {2017})}\BibitemShut
  {NoStop}%
\bibitem [{\citenamefont {Ohnuma}\ \emph
  {et~al.}(2013{\natexlab{b}})\citenamefont {Ohnuma}, \citenamefont {Adachi},
  \citenamefont {Saitoh},\ and\ \citenamefont {Maekawa}}]{ohnuma2013spin}%
  \BibitemOpen
  \bibfield  {author} {\bibinfo {author} {\bibfnamefont {Y.}~\bibnamefont
  {Ohnuma}}, \bibinfo {author} {\bibfnamefont {H.}~\bibnamefont {Adachi}},
  \bibinfo {author} {\bibfnamefont {E.}~\bibnamefont {Saitoh}},\ and\ \bibinfo
  {author} {\bibfnamefont {S.}~\bibnamefont {Maekawa}},\ }\href
  {https://doi.org/https://doi.org/10.1103/PhysRevB.87.014423} {\bibfield
  {journal} {\bibinfo  {journal} {Physical Review B—Condensed Matter and
  Materials Physics}\ }\textbf {\bibinfo {volume} {87}},\ \bibinfo {pages}
  {014423} (\bibinfo {year} {2013}{\natexlab{b}})}\BibitemShut {NoStop}%
\bibitem [{\citenamefont {Holanda}\ \emph {et~al.}(2017)\citenamefont
  {Holanda}, \citenamefont {Alves~Santos}, \citenamefont {Cunha}, \citenamefont
  {Mendes}, \citenamefont {Rodr{\'\i}guez-Su{\'a}rez}, \citenamefont
  {Azevedo},\ and\ \citenamefont {Rezende}}]{holanda2017longitudinal}%
  \BibitemOpen
  \bibfield  {author} {\bibinfo {author} {\bibfnamefont {J.}~\bibnamefont
  {Holanda}}, \bibinfo {author} {\bibfnamefont {O.}~\bibnamefont
  {Alves~Santos}}, \bibinfo {author} {\bibfnamefont {R.}~\bibnamefont {Cunha}},
  \bibinfo {author} {\bibfnamefont {J.}~\bibnamefont {Mendes}}, \bibinfo
  {author} {\bibfnamefont {R.}~\bibnamefont {Rodr{\'\i}guez-Su{\'a}rez}},
  \bibinfo {author} {\bibfnamefont {A.}~\bibnamefont {Azevedo}},\ and\ \bibinfo
  {author} {\bibfnamefont {S.}~\bibnamefont {Rezende}},\ }\href
  {https://doi.org/https://doi.org/10.1103/PhysRevB.95.214421} {\bibfield
  {journal} {\bibinfo  {journal} {Physical Review B}\ }\textbf {\bibinfo
  {volume} {95}},\ \bibinfo {pages} {214421} (\bibinfo {year}
  {2017})}\BibitemShut {NoStop}%
\bibitem [{\citenamefont {Iguchi}\ \emph {et~al.}(2017)\citenamefont {Iguchi},
  \citenamefont {Uchida}, \citenamefont {Daimon},\ and\ \citenamefont
  {Saitoh}}]{PhysRevB.95.174401}%
  \BibitemOpen
  \bibfield  {author} {\bibinfo {author} {\bibfnamefont {R.}~\bibnamefont
  {Iguchi}}, \bibinfo {author} {\bibfnamefont {K.-i.}\ \bibnamefont {Uchida}},
  \bibinfo {author} {\bibfnamefont {S.}~\bibnamefont {Daimon}},\ and\ \bibinfo
  {author} {\bibfnamefont {E.}~\bibnamefont {Saitoh}},\ }\href
  {https://doi.org/10.1103/PhysRevB.95.174401} {\bibfield  {journal} {\bibinfo
  {journal} {Phys. Rev. B}\ }\textbf {\bibinfo {volume} {95}},\ \bibinfo
  {pages} {174401} (\bibinfo {year} {2017})}\BibitemShut {NoStop}%
\bibitem [{\citenamefont {Miura}\ \emph {et~al.}(2017)\citenamefont {Miura},
  \citenamefont {Kikkawa}, \citenamefont {Iguchi}, \citenamefont {Uchida},
  \citenamefont {Saitoh},\ and\ \citenamefont
  {Shiomi}}]{PhysRevMaterials.1.014601}%
  \BibitemOpen
  \bibfield  {author} {\bibinfo {author} {\bibfnamefont {A.}~\bibnamefont
  {Miura}}, \bibinfo {author} {\bibfnamefont {T.}~\bibnamefont {Kikkawa}},
  \bibinfo {author} {\bibfnamefont {R.}~\bibnamefont {Iguchi}}, \bibinfo
  {author} {\bibfnamefont {K.-i.}\ \bibnamefont {Uchida}}, \bibinfo {author}
  {\bibfnamefont {E.}~\bibnamefont {Saitoh}},\ and\ \bibinfo {author}
  {\bibfnamefont {J.}~\bibnamefont {Shiomi}},\ }\href
  {https://doi.org/10.1103/PhysRevMaterials.1.014601} {\bibfield  {journal}
  {\bibinfo  {journal} {Phys. Rev. Mater.}\ }\textbf {\bibinfo {volume} {1}},\
  \bibinfo {pages} {014601} (\bibinfo {year} {2017})}\BibitemShut {NoStop}%
\bibitem [{\citenamefont {Mohmed}\ \emph {et~al.}(2019)\citenamefont {Mohmed},
  \citenamefont {Dar}, \citenamefont {Rubab}, \citenamefont {Hussain},\ and\
  \citenamefont {Hua}}]{mohmed2019magnetic}%
  \BibitemOpen
  \bibfield  {author} {\bibinfo {author} {\bibfnamefont {F.}~\bibnamefont
  {Mohmed}}, \bibinfo {author} {\bibfnamefont {F.~A.}\ \bibnamefont {Dar}},
  \bibinfo {author} {\bibfnamefont {S.}~\bibnamefont {Rubab}}, \bibinfo
  {author} {\bibfnamefont {M.}~\bibnamefont {Hussain}},\ and\ \bibinfo {author}
  {\bibfnamefont {L.-Y.}\ \bibnamefont {Hua}},\ }\href
  {https://doi.org/https://doi.org/10.1016/j.ceramint.2018.10.161} {\bibfield
  {journal} {\bibinfo  {journal} {Ceramics International}\ }\textbf {\bibinfo
  {volume} {45}},\ \bibinfo {pages} {2418} (\bibinfo {year}
  {2019})}\BibitemShut {NoStop}%
\bibitem [{\citenamefont {Czeschka}\ \emph {et~al.}(2011)\citenamefont
  {Czeschka}, \citenamefont {Dreher}, \citenamefont {Brandt}, \citenamefont
  {Weiler}, \citenamefont {Althammer}, \citenamefont {Imort}, \citenamefont
  {Reiss}, \citenamefont {Thomas}, \citenamefont {Schoch}, \citenamefont
  {Limmer}, \citenamefont {Huebl}, \citenamefont {Gross},\ and\ \citenamefont
  {Goennenwein}}]{PhysRevLett.107.046601}%
  \BibitemOpen
  \bibfield  {author} {\bibinfo {author} {\bibfnamefont {F.~D.}\ \bibnamefont
  {Czeschka}}, \bibinfo {author} {\bibfnamefont {L.}~\bibnamefont {Dreher}},
  \bibinfo {author} {\bibfnamefont {M.~S.}\ \bibnamefont {Brandt}}, \bibinfo
  {author} {\bibfnamefont {M.}~\bibnamefont {Weiler}}, \bibinfo {author}
  {\bibfnamefont {M.}~\bibnamefont {Althammer}}, \bibinfo {author}
  {\bibfnamefont {I.-M.}\ \bibnamefont {Imort}}, \bibinfo {author}
  {\bibfnamefont {G.}~\bibnamefont {Reiss}}, \bibinfo {author} {\bibfnamefont
  {A.}~\bibnamefont {Thomas}}, \bibinfo {author} {\bibfnamefont
  {W.}~\bibnamefont {Schoch}}, \bibinfo {author} {\bibfnamefont
  {W.}~\bibnamefont {Limmer}}, \bibinfo {author} {\bibfnamefont
  {H.}~\bibnamefont {Huebl}}, \bibinfo {author} {\bibfnamefont
  {R.}~\bibnamefont {Gross}},\ and\ \bibinfo {author} {\bibfnamefont
  {S.~T.~B.}\ \bibnamefont {Goennenwein}},\ }\href
  {https://doi.org/10.1103/PhysRevLett.107.046601} {\bibfield  {journal}
  {\bibinfo  {journal} {Phys. Rev. Lett.}\ }\textbf {\bibinfo {volume} {107}},\
  \bibinfo {pages} {046601} (\bibinfo {year} {2011})}\BibitemShut {NoStop}%
\bibitem [{\citenamefont {Uchida}\ \emph {et~al.}(2013)\citenamefont {Uchida},
  \citenamefont {Nonaka}, \citenamefont {Kikkawa}, \citenamefont {Kajiwara},\
  and\ \citenamefont {Saitoh}}]{PhysRevB.87.104412}%
  \BibitemOpen
  \bibfield  {author} {\bibinfo {author} {\bibfnamefont {K.}~\bibnamefont
  {Uchida}}, \bibinfo {author} {\bibfnamefont {T.}~\bibnamefont {Nonaka}},
  \bibinfo {author} {\bibfnamefont {T.}~\bibnamefont {Kikkawa}}, \bibinfo
  {author} {\bibfnamefont {Y.}~\bibnamefont {Kajiwara}},\ and\ \bibinfo
  {author} {\bibfnamefont {E.}~\bibnamefont {Saitoh}},\ }\href
  {https://doi.org/10.1103/PhysRevB.87.104412} {\bibfield  {journal} {\bibinfo
  {journal} {Phys. Rev. B}\ }\textbf {\bibinfo {volume} {87}},\ \bibinfo
  {pages} {104412} (\bibinfo {year} {2013})}\BibitemShut {NoStop}%
\bibitem [{\citenamefont {Zhang}\ \emph {et~al.}(2011)\citenamefont {Zhang},
  \citenamefont {Hikino},\ and\ \citenamefont {Yunoki}}]{zhang2011first}%
  \BibitemOpen
  \bibfield  {author} {\bibinfo {author} {\bibfnamefont {Q.}~\bibnamefont
  {Zhang}}, \bibinfo {author} {\bibfnamefont {S.-i.}\ \bibnamefont {Hikino}},\
  and\ \bibinfo {author} {\bibfnamefont {S.}~\bibnamefont {Yunoki}},\ }\href
  {https://pubs.aip.org/aip/apl/article/99/17/172105/341198/First-principles-study-of-the-spin-mixing}
  {\bibfield  {journal} {\bibinfo  {journal} {Applied Physics Letters}\
  }\textbf {\bibinfo {volume} {99}} (\bibinfo {year} {2011})}\BibitemShut
  {NoStop}%
\bibitem [{\citenamefont {Pham}\ \emph {et~al.}(2018)\citenamefont {Pham},
  \citenamefont {Ribeiro}, \citenamefont {Park}, \citenamefont {Lee},
  \citenamefont {Kang}, \citenamefont {Park}, \citenamefont {Nguyen},
  \citenamefont {Michel}, \citenamefont {Yoon}, \citenamefont {Cho} \emph
  {et~al.}}]{pham2018interface}%
  \BibitemOpen
  \bibfield  {author} {\bibinfo {author} {\bibfnamefont {T.~K.~H.}\
  \bibnamefont {Pham}}, \bibinfo {author} {\bibfnamefont {M.}~\bibnamefont
  {Ribeiro}}, \bibinfo {author} {\bibfnamefont {J.~H.}\ \bibnamefont {Park}},
  \bibinfo {author} {\bibfnamefont {N.~J.}\ \bibnamefont {Lee}}, \bibinfo
  {author} {\bibfnamefont {K.~H.}\ \bibnamefont {Kang}}, \bibinfo {author}
  {\bibfnamefont {E.}~\bibnamefont {Park}}, \bibinfo {author} {\bibfnamefont
  {V.~Q.}\ \bibnamefont {Nguyen}}, \bibinfo {author} {\bibfnamefont
  {A.}~\bibnamefont {Michel}}, \bibinfo {author} {\bibfnamefont {C.~S.}\
  \bibnamefont {Yoon}}, \bibinfo {author} {\bibfnamefont {S.}~\bibnamefont
  {Cho}}, \emph {et~al.},\ }\href
  {https://doi.org/https://doi.org/10.1038/s41598-018-31915-3} {\bibfield
  {journal} {\bibinfo  {journal} {Scientific Reports}\ }\textbf {\bibinfo
  {volume} {8}},\ \bibinfo {pages} {13907} (\bibinfo {year}
  {2018})}\BibitemShut {NoStop}%
\bibitem [{\citenamefont {Aqeel}\ \emph {et~al.}(2014)\citenamefont {Aqeel},
  \citenamefont {Vera-Marun}, \citenamefont {van Wees},\ and\ \citenamefont
  {Palstra}}]{aqeel2014surface}%
  \BibitemOpen
  \bibfield  {author} {\bibinfo {author} {\bibfnamefont {A.}~\bibnamefont
  {Aqeel}}, \bibinfo {author} {\bibfnamefont {I.~J.}\ \bibnamefont
  {Vera-Marun}}, \bibinfo {author} {\bibfnamefont {B.~J.}\ \bibnamefont {van
  Wees}},\ and\ \bibinfo {author} {\bibfnamefont {T.}~\bibnamefont {Palstra}},\
  }\href
  {https://pubs.aip.org/aip/jap/article/116/15/153705/281205/Surface-sensitivity-of-the-spin-Seebeck-effect}
  {\bibfield  {journal} {\bibinfo  {journal} {Journal of Applied Physics}\
  }\textbf {\bibinfo {volume} {116}} (\bibinfo {year} {2014})}\BibitemShut
  {NoStop}%
\bibitem [{\citenamefont {Qiu}\ \emph {et~al.}(2015)\citenamefont {Qiu},
  \citenamefont {Hou}, \citenamefont {Uchida},\ and\ \citenamefont
  {Saitoh}}]{qiu2015influence}%
  \BibitemOpen
  \bibfield  {author} {\bibinfo {author} {\bibfnamefont {Z.}~\bibnamefont
  {Qiu}}, \bibinfo {author} {\bibfnamefont {D.}~\bibnamefont {Hou}}, \bibinfo
  {author} {\bibfnamefont {K.}~\bibnamefont {Uchida}},\ and\ \bibinfo {author}
  {\bibfnamefont {E.}~\bibnamefont {Saitoh}},\ }\href
  {https://doi.org/10.1088/0022-3727/48/16/164013} {\bibfield  {journal}
  {\bibinfo  {journal} {Journal of Physics D: Applied Physics}\ }\textbf
  {\bibinfo {volume} {48}},\ \bibinfo {pages} {164013} (\bibinfo {year}
  {2015})}\BibitemShut {NoStop}%
\bibitem [{\citenamefont {Kim}\ \emph {et~al.}(2020)\citenamefont {Kim},
  \citenamefont {Park},\ and\ \citenamefont {Jin}}]{kim2020enhancing}%
  \BibitemOpen
  \bibfield  {author} {\bibinfo {author} {\bibfnamefont {M.}~\bibnamefont
  {Kim}}, \bibinfo {author} {\bibfnamefont {S.~J.}\ \bibnamefont {Park}},\ and\
  \bibinfo {author} {\bibfnamefont {H.}~\bibnamefont {Jin}},\ }\href
  {https://pubs.aip.org/aip/jap/article/127/8/085105/157180/Enhancing-the-spin-Seebeck-effect-by-controlling}
  {\bibfield  {journal} {\bibinfo  {journal} {Journal of Applied Physics}\
  }\textbf {\bibinfo {volume} {127}} (\bibinfo {year} {2020})}\BibitemShut
  {NoStop}%
\bibitem [{\citenamefont {Kumawat}\ \emph {et~al.}(2024)\citenamefont
  {Kumawat}, \citenamefont {Jain},\ and\ \citenamefont
  {Yusuf}}]{kumawat2024enhanced}%
  \BibitemOpen
  \bibfield  {author} {\bibinfo {author} {\bibfnamefont {K.}~\bibnamefont
  {Kumawat}}, \bibinfo {author} {\bibfnamefont {A.}~\bibnamefont {Jain}},\ and\
  \bibinfo {author} {\bibfnamefont {S.}~\bibnamefont {Yusuf}},\ }\href
  {https://doi.org/https://doi.org/10.1016/j.jallcom.2024.175187} {\bibfield
  {journal} {\bibinfo  {journal} {Journal of Alloys and Compounds}\ }\textbf
  {\bibinfo {volume} {1001}},\ \bibinfo {pages} {175187} (\bibinfo {year}
  {2024})}\BibitemShut {NoStop}%
\bibitem [{\citenamefont {Chang}\ \emph {et~al.}(2017)\citenamefont {Chang},
  \citenamefont {Praveen~Janantha}, \citenamefont {Ding}, \citenamefont {Liu},
  \citenamefont {Cline}, \citenamefont {Gelfand}, \citenamefont {Li},
  \citenamefont {Marconi},\ and\ \citenamefont {Wu}}]{chang2017role}%
  \BibitemOpen
  \bibfield  {author} {\bibinfo {author} {\bibfnamefont {H.}~\bibnamefont
  {Chang}}, \bibinfo {author} {\bibfnamefont {P.}~\bibnamefont
  {Praveen~Janantha}}, \bibinfo {author} {\bibfnamefont {J.}~\bibnamefont
  {Ding}}, \bibinfo {author} {\bibfnamefont {T.}~\bibnamefont {Liu}}, \bibinfo
  {author} {\bibfnamefont {K.}~\bibnamefont {Cline}}, \bibinfo {author}
  {\bibfnamefont {J.~N.}\ \bibnamefont {Gelfand}}, \bibinfo {author}
  {\bibfnamefont {W.}~\bibnamefont {Li}}, \bibinfo {author} {\bibfnamefont
  {M.~C.}\ \bibnamefont {Marconi}},\ and\ \bibinfo {author} {\bibfnamefont
  {M.}~\bibnamefont {Wu}},\ }\href
  {https://doi.org/https://doi.org/10.1126/sciadv.1601614} {\bibfield
  {journal} {\bibinfo  {journal} {Science advances}\ }\textbf {\bibinfo
  {volume} {3}},\ \bibinfo {pages} {e1601614} (\bibinfo {year}
  {2017})}\BibitemShut {NoStop}%
\bibitem [{\citenamefont {Hoffman}\ \emph {et~al.}(2013)\citenamefont
  {Hoffman}, \citenamefont {Sato},\ and\ \citenamefont
  {Tserkovnyak}}]{hoffman2013landau}%
  \BibitemOpen
  \bibfield  {author} {\bibinfo {author} {\bibfnamefont {S.}~\bibnamefont
  {Hoffman}}, \bibinfo {author} {\bibfnamefont {K.}~\bibnamefont {Sato}},\ and\
  \bibinfo {author} {\bibfnamefont {Y.}~\bibnamefont {Tserkovnyak}},\ }\href
  {https://doi.org/https://doi.org/10.1103/PhysRevB.88.064408} {\bibfield
  {journal} {\bibinfo  {journal} {Physical Review B—Condensed Matter and
  Materials Physics}\ }\textbf {\bibinfo {volume} {88}},\ \bibinfo {pages}
  {064408} (\bibinfo {year} {2013})}\BibitemShut {NoStop}%
\bibitem [{\citenamefont {Li}\ \emph {et~al.}(2022)\citenamefont {Li},
  \citenamefont {Zheng}, \citenamefont {Fang}, \citenamefont {Liu},
  \citenamefont {Zhang}, \citenamefont {Chen}, \citenamefont {Ma},
  \citenamefont {Shen}, \citenamefont {Liu}, \citenamefont {Manchon} \emph
  {et~al.}}]{li2022unconventional}%
  \BibitemOpen
  \bibfield  {author} {\bibinfo {author} {\bibfnamefont {Y.}~\bibnamefont
  {Li}}, \bibinfo {author} {\bibfnamefont {D.}~\bibnamefont {Zheng}}, \bibinfo
  {author} {\bibfnamefont {B.}~\bibnamefont {Fang}}, \bibinfo {author}
  {\bibfnamefont {C.}~\bibnamefont {Liu}}, \bibinfo {author} {\bibfnamefont
  {C.}~\bibnamefont {Zhang}}, \bibinfo {author} {\bibfnamefont
  {A.}~\bibnamefont {Chen}}, \bibinfo {author} {\bibfnamefont {Y.}~\bibnamefont
  {Ma}}, \bibinfo {author} {\bibfnamefont {K.}~\bibnamefont {Shen}}, \bibinfo
  {author} {\bibfnamefont {H.}~\bibnamefont {Liu}}, \bibinfo {author}
  {\bibfnamefont {A.}~\bibnamefont {Manchon}}, \emph {et~al.},\ }\href
  {https://doi.org/https://doi.org/10.1002/adma.202200019} {\bibfield
  {journal} {\bibinfo  {journal} {Advanced Materials}\ }\textbf {\bibinfo
  {volume} {34}},\ \bibinfo {pages} {2200019} (\bibinfo {year}
  {2022})}\BibitemShut {NoStop}%
\bibitem [{\citenamefont {Kalarickal}\ \emph {et~al.}(2009)\citenamefont
  {Kalarickal}, \citenamefont {Mo}, \citenamefont {Krivosik},\ and\
  \citenamefont {Patton}}]{kalarickal2009ferromagnetic}%
  \BibitemOpen
  \bibfield  {author} {\bibinfo {author} {\bibfnamefont {S.~S.}\ \bibnamefont
  {Kalarickal}}, \bibinfo {author} {\bibfnamefont {N.}~\bibnamefont {Mo}},
  \bibinfo {author} {\bibfnamefont {P.}~\bibnamefont {Krivosik}},\ and\
  \bibinfo {author} {\bibfnamefont {C.~E.}\ \bibnamefont {Patton}},\ }\href
  {https://doi.org/https://doi.org/10.1103/PhysRevB.79.094427} {\bibfield
  {journal} {\bibinfo  {journal} {Physical Review B—Condensed Matter and
  Materials Physics}\ }\textbf {\bibinfo {volume} {79}},\ \bibinfo {pages}
  {094427} (\bibinfo {year} {2009})}\BibitemShut {NoStop}%
\bibitem [{\citenamefont {R{\"o}schmann}\ and\ \citenamefont
  {Winkler}(1977)}]{roschmann1977relaxation}%
  \BibitemOpen
  \bibfield  {author} {\bibinfo {author} {\bibfnamefont {P.}~\bibnamefont
  {R{\"o}schmann}}\ and\ \bibinfo {author} {\bibfnamefont {G.}~\bibnamefont
  {Winkler}},\ }\href
  {https://doi.org/https://doi.org/10.1016/0304-8853(77)90021-X} {\bibfield
  {journal} {\bibinfo  {journal} {Journal of Magnetism and Magnetic Materials}\
  }\textbf {\bibinfo {volume} {4}},\ \bibinfo {pages} {105} (\bibinfo {year}
  {1977})}\BibitemShut {NoStop}%
\bibitem [{\citenamefont {R{\"o}schmann}(1976)}]{roschmann1976two}%
  \BibitemOpen
  \bibfield  {author} {\bibinfo {author} {\bibfnamefont {P.}~\bibnamefont
  {R{\"o}schmann}},\ }in\ \href
  {https://doi.org/https://doi.org/10.1063/1.2946091} {\emph {\bibinfo
  {booktitle} {AIP Conference Proceedings}}},\ Vol.~\bibinfo {volume} {34}\
  (\bibinfo {organization} {American Institute of Physics},\ \bibinfo {year}
  {1976})\ pp.\ \bibinfo {pages} {253--258}\BibitemShut {NoStop}%
\bibitem [{\citenamefont {Nguyen}\ \emph {et~al.}(2017)\citenamefont {Nguyen},
  \citenamefont {Sandilands}, \citenamefont {Sohn}, \citenamefont {Kim},
  \citenamefont {Wysocki}, \citenamefont {Yang}, \citenamefont {Moon},
  \citenamefont {Ko}, \citenamefont {Yamaura}, \citenamefont {Hiroi} \emph
  {et~al.}}]{nguyen2017two}%
  \BibitemOpen
  \bibfield  {author} {\bibinfo {author} {\bibfnamefont {T.~M.~H.}\
  \bibnamefont {Nguyen}}, \bibinfo {author} {\bibfnamefont {L.~J.}\
  \bibnamefont {Sandilands}}, \bibinfo {author} {\bibfnamefont
  {C.}~\bibnamefont {Sohn}}, \bibinfo {author} {\bibfnamefont {C.~H.}\
  \bibnamefont {Kim}}, \bibinfo {author} {\bibfnamefont {A.~L.}\ \bibnamefont
  {Wysocki}}, \bibinfo {author} {\bibfnamefont {I.-S.}\ \bibnamefont {Yang}},
  \bibinfo {author} {\bibfnamefont {S.}~\bibnamefont {Moon}}, \bibinfo {author}
  {\bibfnamefont {J.-H.}\ \bibnamefont {Ko}}, \bibinfo {author} {\bibfnamefont
  {J.}~\bibnamefont {Yamaura}}, \bibinfo {author} {\bibfnamefont
  {Z.}~\bibnamefont {Hiroi}}, \emph {et~al.},\ }\href
  {https://doi.org/https://doi.org/10.1038/s41467-017-00228-w} {\bibfield
  {journal} {\bibinfo  {journal} {Nature communications}\ }\textbf {\bibinfo
  {volume} {8}},\ \bibinfo {pages} {251} (\bibinfo {year} {2017})}\BibitemShut
  {NoStop}%
\bibitem [{\citenamefont {Hurben}\ and\ \citenamefont
  {Patton}(1998)}]{hurben1998theory}%
  \BibitemOpen
  \bibfield  {author} {\bibinfo {author} {\bibfnamefont {M.}~\bibnamefont
  {Hurben}}\ and\ \bibinfo {author} {\bibfnamefont {C.}~\bibnamefont
  {Patton}},\ }\href {https://doi.org/https://doi.org/10.1063/1.367194}
  {\bibfield  {journal} {\bibinfo  {journal} {Journal of Applied Physics}\
  }\textbf {\bibinfo {volume} {83}},\ \bibinfo {pages} {4344} (\bibinfo {year}
  {1998})}\BibitemShut {NoStop}%
\bibitem [{\citenamefont {Li}\ \emph {et~al.}(2024)\citenamefont {Li},
  \citenamefont {Zhang}, \citenamefont {Liu}, \citenamefont {Zheng},
  \citenamefont {Fang}, \citenamefont {Zhang}, \citenamefont {Chen},
  \citenamefont {Ma}, \citenamefont {Wang}, \citenamefont {Liu} \emph
  {et~al.}}]{li2024reconfigurable}%
  \BibitemOpen
  \bibfield  {author} {\bibinfo {author} {\bibfnamefont {Y.}~\bibnamefont
  {Li}}, \bibinfo {author} {\bibfnamefont {Z.}~\bibnamefont {Zhang}}, \bibinfo
  {author} {\bibfnamefont {C.}~\bibnamefont {Liu}}, \bibinfo {author}
  {\bibfnamefont {D.}~\bibnamefont {Zheng}}, \bibinfo {author} {\bibfnamefont
  {B.}~\bibnamefont {Fang}}, \bibinfo {author} {\bibfnamefont {C.}~\bibnamefont
  {Zhang}}, \bibinfo {author} {\bibfnamefont {A.}~\bibnamefont {Chen}},
  \bibinfo {author} {\bibfnamefont {Y.}~\bibnamefont {Ma}}, \bibinfo {author}
  {\bibfnamefont {C.}~\bibnamefont {Wang}}, \bibinfo {author} {\bibfnamefont
  {H.}~\bibnamefont {Liu}}, \emph {et~al.},\ }\href
  {https://doi.org/https://doi.org/10.1038/s41467-024-46330-8} {\bibfield
  {journal} {\bibinfo  {journal} {Nature Communications}\ }\textbf {\bibinfo
  {volume} {15}},\ \bibinfo {pages} {2234} (\bibinfo {year}
  {2024})}\BibitemShut {NoStop}%
\bibitem [{\citenamefont {Wang}\ \emph {et~al.}(2020)\citenamefont {Wang},
  \citenamefont {Xie}, \citenamefont {Xu},\ and\ \citenamefont
  {Xia}}]{PhysRevB.101.165137}%
  \BibitemOpen
  \bibfield  {author} {\bibinfo {author} {\bibfnamefont {L.-W.}\ \bibnamefont
  {Wang}}, \bibinfo {author} {\bibfnamefont {L.-S.}\ \bibnamefont {Xie}},
  \bibinfo {author} {\bibfnamefont {P.-X.}\ \bibnamefont {Xu}},\ and\ \bibinfo
  {author} {\bibfnamefont {K.}~\bibnamefont {Xia}},\ }\href
  {https://doi.org/10.1103/PhysRevB.101.165137} {\bibfield  {journal} {\bibinfo
   {journal} {Phys. Rev. B}\ }\textbf {\bibinfo {volume} {101}},\ \bibinfo
  {pages} {165137} (\bibinfo {year} {2020})}\BibitemShut {NoStop}%
\bibitem [{\citenamefont {Harris}(1963)}]{harris1963spin}%
  \BibitemOpen
  \bibfield  {author} {\bibinfo {author} {\bibfnamefont {A.~B.}\ \bibnamefont
  {Harris}},\ }\href {https://doi.org/https://doi.org/10.1103/PhysRev.132.2398}
  {\bibfield  {journal} {\bibinfo  {journal} {Physical Review}\ }\textbf
  {\bibinfo {volume} {132}},\ \bibinfo {pages} {2398} (\bibinfo {year}
  {1963})}\BibitemShut {NoStop}%
\bibitem [{\citenamefont {Vasili}\ \emph {et~al.}(2017)\citenamefont {Vasili},
  \citenamefont {Casals}, \citenamefont {Cichelero}, \citenamefont {Maci{\`a}},
  \citenamefont {Geshev}, \citenamefont {Gargiani}, \citenamefont {Valvidares},
  \citenamefont {Herrero-Martin}, \citenamefont {Pellegrin}, \citenamefont
  {Fontcuberta} \emph {et~al.}}]{vasili2017direct}%
  \BibitemOpen
  \bibfield  {author} {\bibinfo {author} {\bibfnamefont {H.~B.}\ \bibnamefont
  {Vasili}}, \bibinfo {author} {\bibfnamefont {B.}~\bibnamefont {Casals}},
  \bibinfo {author} {\bibfnamefont {R.}~\bibnamefont {Cichelero}}, \bibinfo
  {author} {\bibfnamefont {F.}~\bibnamefont {Maci{\`a}}}, \bibinfo {author}
  {\bibfnamefont {J.}~\bibnamefont {Geshev}}, \bibinfo {author} {\bibfnamefont
  {P.}~\bibnamefont {Gargiani}}, \bibinfo {author} {\bibfnamefont
  {M.}~\bibnamefont {Valvidares}}, \bibinfo {author} {\bibfnamefont
  {J.}~\bibnamefont {Herrero-Martin}}, \bibinfo {author} {\bibfnamefont
  {E.}~\bibnamefont {Pellegrin}}, \bibinfo {author} {\bibfnamefont
  {J.}~\bibnamefont {Fontcuberta}}, \emph {et~al.},\ }\href
  {https://doi.org/https://doi.org/10.1103/PhysRevB.96.014433} {\bibfield
  {journal} {\bibinfo  {journal} {Physical Review B}\ }\textbf {\bibinfo
  {volume} {96}},\ \bibinfo {pages} {014433} (\bibinfo {year}
  {2017})}\BibitemShut {NoStop}%
\bibitem [{\citenamefont {Imamura}\ \emph {et~al.}(2021)\citenamefont
  {Imamura}, \citenamefont {Asada}, \citenamefont {Nishimura}, \citenamefont
  {Yamaguchi}, \citenamefont {Tashima},\ and\ \citenamefont
  {Kitagawa}}]{imamura2021enhancement}%
  \BibitemOpen
  \bibfield  {author} {\bibinfo {author} {\bibfnamefont {M.}~\bibnamefont
  {Imamura}}, \bibinfo {author} {\bibfnamefont {H.}~\bibnamefont {Asada}},
  \bibinfo {author} {\bibfnamefont {R.}~\bibnamefont {Nishimura}}, \bibinfo
  {author} {\bibfnamefont {K.}~\bibnamefont {Yamaguchi}}, \bibinfo {author}
  {\bibfnamefont {D.}~\bibnamefont {Tashima}},\ and\ \bibinfo {author}
  {\bibfnamefont {J.}~\bibnamefont {Kitagawa}},\ }\href
  {https://pubs.aip.org/aip/adv/article/11/3/035143/991083/Enhancement-of-the-spin-Seebeck-effect-owing-to}
  {\bibfield  {journal} {\bibinfo  {journal} {AIP Advances}\ }\textbf {\bibinfo
  {volume} {11}} (\bibinfo {year} {2021})}\BibitemShut {NoStop}%
\bibitem [{\citenamefont {Xie}\ \emph {et~al.}(2017)\citenamefont {Xie},
  \citenamefont {Jin}, \citenamefont {He}, \citenamefont {Bauer}, \citenamefont
  {Barker},\ and\ \citenamefont {Xia}}]{xie2017first}%
  \BibitemOpen
  \bibfield  {author} {\bibinfo {author} {\bibfnamefont {L.-S.}\ \bibnamefont
  {Xie}}, \bibinfo {author} {\bibfnamefont {G.-X.}\ \bibnamefont {Jin}},
  \bibinfo {author} {\bibfnamefont {L.}~\bibnamefont {He}}, \bibinfo {author}
  {\bibfnamefont {G.~E.}\ \bibnamefont {Bauer}}, \bibinfo {author}
  {\bibfnamefont {J.}~\bibnamefont {Barker}},\ and\ \bibinfo {author}
  {\bibfnamefont {K.}~\bibnamefont {Xia}},\ }\href
  {https://doi.org/https://doi.org/10.1103/PhysRevB.95.014423} {\bibfield
  {journal} {\bibinfo  {journal} {Physical Review B}\ }\textbf {\bibinfo
  {volume} {95}},\ \bibinfo {pages} {014423} (\bibinfo {year}
  {2017})}\BibitemShut {NoStop}%
\end{thebibliography}%

\end{document}